\documentclass[aps,prc,twocolumn,floatfix,superscriptaddress,nofootinbib]{revtex4-1}

\usepackage{graphicx}
\usepackage{bm}
\usepackage{physics}
\usepackage{longtable}
\usepackage[colorlinks,citecolor=blue,bookmarks]{hyperref}
\usepackage{makecell}
\newcolumntype{T}[1]{>{\centering\arraybackslash}p{#1}}

\newcommand{\elma}[3]{\bra{#1} #2 \ket{#3}}

\begin{document}

\title{M1 dipole strength from projected generator coordinate method calculations in the $sd$-shell valence space}

\author{S. Bofos}
\affiliation{CEA, DES, IRESNE, DER, SPRC, LEPh, France} 

\author{J. Mart\'inez-Larraz} 
\affiliation{Departamento de F\'isica Te\'orica, Universidad Aut\'onoma de Madrid, E-28049 Madrid, Spain}

\author{B. Bally}
\affiliation{Technische Universit\"at Darmstadt, Department of Physics, 64289 Darmstadt, Germany}
\affiliation{ESNT, IRFU, CEA, Universit\'e Paris-Saclay, 91191 Gif-sur-Yvette, France}

\author{T. Duguet}
\affiliation{IRFU, CEA, Universit\'e Paris-Saclay, 91191 Gif-sur-Yvette, France}  
\affiliation{KU Leuven, Department of Physics and Astronomy, Instituut voor Kern- en Stralingsfysica, 3001 Leuven, Belgium}  

\author{M. Frosini}
\affiliation{CEA, DES, IRESNE, DER, SPRC, LEPh, France}

\author{T. R. Rodr\'iguez}
\affiliation{Departamento de F\'isica At\'omica, Molecular y Nuclear, Universidad de Sevilla, E-41012 Sevilla, Spain}
\affiliation{Grupo de F\'isica Nuclear, Departamento EMFTEL and IPARCOS, Universidad Complutense de Madrid, E-28040 Madrid, Spain}

\author{K. Sieja}
\affiliation{Universit\'e de Strasbourg, IPHC, 23 rue du Loess 67037 Strasbourg, France
CNRS, UMR7178, 67037 Strasbourg, France}
  
\date{\today} 

\begin{abstract}
\begin{description}
\item[Background] 
The low-energy enhancement observed in the deexcitation $\gamma$-ray strength functions,  attributed to magnetic dipole ($M1$) radiations, has spurred theoretical efforts to improve on its description. Among the most widely used approaches are the quasiparticle random-phase approximation (QRPA) and its extensions. However, these methods often struggle to reproduce the correct behavior of the $M1$ strength at the lowest $\gamma$ energies.
An alternative framework, the projected generator coordinate method (PGCM), offers significant advantages over QRPA by restoring broken symmetries and incorporating both vibrational and rotational dynamics within a unified description. 
Due to these features, PGCM has been proposed as a promising tool to study the low-energy $M1$ strength function in atomic nuclei. However, comprehensive investigations employing this method are lacking.
\item[Purpose] The PGCM is developed and tested in its description of $1^+$ excited states in even-even nuclei, along with the $M1$ transitions connecting them to the $0^+$ ground state.
\item[Methods]
The PGCM  is presently used within the frame of $sd$-shell valence space calculations based on the USDB shell-model interaction to benchmark its performance against the solutions obtained via exact diagonalization, i.e. results from the latter are considered as ``exact'' reference results in the present study. The reliability of two different sets of generator coordinates in the PGCM calculations is gauged using $^{24}$Mg as a test case.
\item[Results]
Energies and $M1$ transition strengths of the lowest excited $1^+$ states extracted from the PGCM calculation reproduce well the exact results for both sets of collective coordinates. Eventually, PGCM accurately models the exact level density and cumulative $B(M1)$ strength up to about $20$~MeV excitation energy in all considered nuclei.
\item[Conclusions] The ability of the PGCM to reproduce results from exact diagonalization in the $sd$ valence space is demonstrated for $1^+$ states and $M1$ transitions. Future work will need to assess whether the proposed method can be applied systematically and extended to large-scale calculations while maintaining a reasonable computational cost.
\end{description}
\end{abstract}

\maketitle

Electromagnetic transition properties are an essential ingredient in the evaluation of neutron inelastic and capture cross sections
at play in astrophysical and reactor applications. Even if all multipolarities may contribute in theory, 
electromagnetic transition rates are largely dominated by dipole modes, namely, the electric giant dipole resonance (GDR) 
at high energy and the magnetic spin-flip transitions at low energy. Traditionally, 
dipole strength functions have been modeled using phenomenological Lorentzian approximations \cite{SLO,CAPOTE20093107,RIPL-3}. 
However, important deviations from statistical behavior were evidenced at low $\gamma$-ray energies, especially the so-called low-energy enhancement (LEE) of the dipole 
strength~\cite{Larsen-Fe56,enh-Mo,end-Sc43,Larsen-Ti44,Larsen2010,oslo-website}, 
motivating the evaluation of strength functions on more microscopic grounds.
The two aforementioned dipole modes have been extensively studied in nuclear structure calculations, mainly within the quasiparticle random phase approximation (QRPA) \cite{RingSchuck,Paar2006,Martini2016,Goriely2016}.
This method has proven useful
to generate strength functions for large-scale evaluation of cross sections, see, e.g.,~\cite{Goriely2018}. 
However, the QRPA formalism suffers from several limitations. These include the lack of 
correlations due to the underlying harmonic approximation, incomplete restoration of symmetries~\cite{Porro:2023yto}, the quasi-boson approximation, which introduces a violation of the Pauli exclusion principle,
and the truncation of the many-body Hilbert space to two-quasiparticle excitations on top of the reference mean-field states. 
Recently, those shortcomings were put in evidence in a theoretical benchmark study of $M1$ transitions in valence space calculations~\cite{Frosini2023}, where the finite-temperature QRPA method appeared insufficient to provide a correct description of the LEE when compared to  results of the exact shell-model (SM) diagonalization. 

In this work, the projected generator coordinate method (PGCM) is proposed
as an alternative solution to overcome the limitations of the QRPA. 
The advantage of the PGCM lies in the improved wave functions generated by
symmetry restoration, the absence of a harmonic approximation, and the incorporation of a well-tailored collective degrees of freedom. 
The method has been routinely used to describe low-lying collective states 
and electromagnetic properties, both with energy density functionals and Hamiltonians 
(see, e.g., Refs.~\cite{Bender08,Niksic11,Robledo19,Gao15,Jiao17,Bally19,Sanchez2021a,Dao22} and references therein). 
The structure of the dipole resonances (GDR and also pygmy dipole resonance, PDR)
was successfully attempted in the angular-momentum projected GCM approach with basis 
functions produced by means of shifted-basis antisymmetrized molecular dynamics \cite{Kimura2017}.   
Recently, the PGCM has been employed successfully within the frame of ab-initio calculations. While reproducing absolute energies necessitates the addition of dynamical correlations on top of it~\cite{Frosini:2021fjf,Frosini:2021sxj,Frosini:2021ddm}, the PGCM was shown to provide an accurate description of the low-lying collective states as well as of the main characteristics of the giant monopole resonance in mid-mass nuclei~\cite{Porro:2024vlc,Porro:2024tzt,Porro:2024pdn,Porro:2024bid}.

The goal of the present work is to gauge the capacity of the PGCM to describe $M1$ 
transitions by benchmarking the results obtained within the frame of valence space calculations against those generated via exact diagonalization. 
The present study first focuses on $^{24}$Mg, one of the nuclei examined in Ref.~\cite{Frosini2023}, by addressing the description of $1^+$ excited states that require breaking both rotational and time-reversal symmetries in PGCM calculations. The magnetic moments of those states and the $M1$ transitions between them as well as to the $0^+$ ground state are investigated. Finally, the capacity of the PGCM to reproduce the $B(M1)$ strength in a number of the $sd$-shell nuclei is demonstrated. 

The paper is organized as follows. 
First, Sec.~\ref{sec:theory} introduces the theoretical ingredients at play in the present study.
Then, PGCM results obtained in ${}^{24}$Mg are benchmarked in detail against results obtained from the SM diagonalization in Sec.~\ref{sec:results} . The comparison is further extended to other $sd$-shell nuclei in Sec.~\ref{sec:results:other}.
Finally, Sec.~\ref{sec:conclusion} provides the conclusions and perspectives of the present work.

\section{Theoretical framework}
\label{sec:theory}

\subsection{Model space and Hamiltonian}
\label{sec:theory:model}

Calculations are presently performed using the $sd$-shell valence space consisting of the $0d_{5/2}$, $1s_{1/2}$ and $0d_{3/2}$ spherical harmonic orbitals for both protons and neutrons on top of an ${}^{16}$O inert core. 
For the nuclear Hamiltonian $H$, the USDB effective and phenomenological two-body interaction \cite{USDB} is employed.
While the primary focus of this article is the benchmark comparison between two theoretical schemes, it must be noted that the exact diagonalization based on the USDB interaction provides an excellent description of spectroscopy and of experimental magnetic dipole moments, $M1$ transitions and $B(\mathrm{GT})$ values in this region when applying standard quenching factors to the corresponding operators~\cite{USDB,Richter2008,Richter2009}.

\subsection{Schrödinger equation}
\label{sec:theory:exact}

Within the restricted many-body Hilbert space built out of the one-body $sd$ valence space, the structure of nuclear systems is governed by the many-body Schrödinger equation
\begin{equation}
\label{eq:schrodinger}
    H \ket{\Psi^{\Lambda M}_\sigma} = E^{\Lambda}_\sigma \ket{\Psi^{\Lambda M}_\sigma} ,
\end{equation}
where $\ket{\Psi^{\Lambda M}_\sigma}$ denotes an eigenstate of $H$ with eigenenergy $E^{\Lambda}_\sigma$, 
characterized by symmetry quantum numbers: the numbers of valence protons $Z_v$ and neutrons $N_v$, the total angular momentum $J$ and its third component $M$, and the parity $\pi$. These quantum numbers are collectively denoted as $\Lambda M \equiv Z_v N_v J \pi M$. The total proton $Z$ and neutron $N$ numbers are obtained by adding the core contributions, which, for the $sd$-shell gives $Z = Z_v + 8$ and $N = N_v + 8$.
The index $\sigma = 1, 2, \ldots$ distinguishes different eigenstates with the same $\Lambda$ values.

The $sd$-shell valence space is sufficiently small to allow for an exact solving of Eq.~\eqref{eq:schrodinger} through the diagonalization of the Hamiltonian matrix. 
The $j$-coupled shell-model code NATHAN~\cite{RMP} is presently employed to obtain exact eigenvectors and eigenvalues of $H$ for the nuclei of interest. The corresponding results are hereafter dubbed as ``exact'' or shell-model results.  

\subsection{Observables related to the $M1$ operator}
\label{sec:theory:M1}

In this work, the usual one-body magnetic operator $M^{1 \mu}$ ($\mu = 0, \pm 1 $) is employed, i.e.,
\begin{equation}
    \label{eq: M1}
    M^{1 \mu} \equiv \sqrt{\frac{3}{4\pi}} \mu_N \sum_{i=p,n} \left( g_{i,l} \, l_{i,\mu} + g_{i,s} \, s_{i,\mu} \right) ,
\end{equation}
where $g_{i,l}$ denotes the orbital g-factor, $l_{i,\mu}$ the orbital angular momentum operator, $g_{i,s}$ the spin g-factor and $s_{i,\mu}$ the spin operator associated with protons ($i=p$) and neutrons ($i=n$), while $\mu_N$ is the nuclear magneton. Bare orbital g-factors are used whereas spin g-factors are multiplied by 0.75 as is customary in valence space calculations, see e.g.~Ref.~\cite{RMP} and references therein. In any case, the particular choice made for the g factors is of minor
importance for this study whose primary aim is to compare two different many-body methods and not to reproduce experimental data.

The matrix elements of the operators $M^{1 \mu}$ are associated with several observables of present interest. 
First, the magnetic dipole moment of $J^\pi_\sigma = 1_\sigma^+$ states is given by
\begin{equation}
\mu(1_\sigma^+) \equiv \sqrt{\frac{4\pi}{3}} \elma{\Psi_\sigma^{\Lambda_i 1}}{M^{10}}{\Psi_\sigma^{\Lambda_i 1}} \, ,
\end{equation}
with $\Lambda_i \equiv Z_v N_v 1 1$. Furthermore, the reduced transition probability from an excited  $1_\sigma^+$ state to the $0^+_1$ ground state reads as
\begin{equation}
\begin{split}
B(M1; 1_\sigma^+ \rightarrow 0^+_1) \equiv  \frac{1}{3} \sum_{\mu = -1}^1 \sum_{M=-1}^1  \vert \langle \Psi_1^{\Lambda_f 0} \vert M^{1 \mu} \vert \Psi_\sigma^{\Lambda_i M} \rangle \vert^2  \, ,
\end{split}
\end{equation}
where $\Lambda_f \equiv  Z_v N_v 0 1$.
It is eventually customary to fold the $B(M1)$ distributions with a combination of Lorentzians of given width $\Gamma$ (see, e.g., Figs.~\ref{fig:pgcm_sum_lorentz_mu} and \ref{fig:other_nuclei} with $\Gamma = 0.5$ MeV):
\begin{equation}
    \label{eq:convolution_funct}
\begin{aligned}
    B(M1)(\omega) = &\frac{1}{\pi} \sum_{\sigma=1}^{\sigma_{\mathrm{max}}} B(M1; 1_\sigma^+ \rightarrow 0^+_1) \frac{\Gamma}{(E(1^+_\sigma) - \omega)^2 + \Gamma^2} \\
    -&\frac{1}{\pi} \sum_{\sigma=1}^{\sigma_{\mathrm{max}}} B(M1; 1_\sigma^+ \rightarrow 0^+_1) \frac{\Gamma}{(E(1^+_\sigma)  + \omega)^2 + \Gamma^2}.
\end{aligned}
\end{equation}

\subsection{Projected generator coordinate method}
\label{sec:theory:pgcm}

The PGCM is an approximate method based on the variational space spanned by a set of non-orthogonal symmetry-projected Bogoliubov states~\cite{RingSchuck} 
\begin{equation}
  \label{eq:pgcm}
  \ket{\Theta^{\Lambda M}_\sigma} \equiv \sum_{qK} f^{\Lambda}_{\sigma; K} (q) P^{\Lambda}_{M K} \ket{\Phi(q)} ,  
\end{equation}  
where \(q\) denotes the so-called (possibly multidimensional) generator coordinate(s) whose specific value(s) characterize the seed states obtained by solving constrained Hartree-Fock-Bogoliubov (HFB) equations (see below).
The projection operators $P^{\Lambda}_{M K} \equiv P^Z P^N P^J_{M K}$ select the component with the desired quantum numbers associated with the symmetries of $H$.
Finally, the mixing coefficients $f^{\Lambda}_{\sigma;K} (q)$ are determined based on Ritz' variational principle, which leads to diagonalizing $H$ within the subspace spanned by the set of non-orthogonal projected states by solving the Hill-Wheeler-Griffin (HWG) equation
\begin{equation}
   \label{eq:hwg}
    \sum\limits_{qK} H^{\Lambda}_{q'K'qK} f_{\sigma; K}^{\Lambda}\qty(q) =
    E_{\sigma}^{\Lambda} \sum\limits_{qK} N^{\Lambda}_{q'K'qK} f_{\sigma; K}^{\Lambda}\qty(q)\,,
\end{equation}
where 
\begin{subequations}
\begin{align}
   \label{eq:norm_kernels}
    N^{\Lambda}_{q'K'qK} &\equiv \langle \Phi(q') | P^{\Lambda}_{K'K} |\Phi(q) \rangle , \\
    \label{eq:hamiltonian_kernels}
    H^{\Lambda}_{q'K'qK} &\equiv \langle \Phi(q') | H P^{\Lambda}_{K'K}  |\Phi(q) \rangle
\end{align}
\end{subequations}
are the projected norm and Hamiltonian kernels, respectively. 

A key aspect of the PGCM is the selection of the mean-field states that, after the symmetry restoration, define the variational subspace.
The ability of these states to efficiently explore the Hilbert space associated with the observables under study is crucial for the quality of the calculation. This ability depends to a large extent on the symmetries that the mean-field states are allowed to break.

While Bogoliubov states obtained by solving variational HFB equations can spontaneously break symmetries, employing a variation after projection (VAP) technique offers the possibility to better optimize the degree of symmetry breaking~\cite{RingSchuck}. While VAP can be applied to all symmetries in principle, the variation after angular momentum projection (VAAMP) is rarely used in practice due to its prohibitive computational cost in realistic scenarios, with only a few implementations reported to date~\cite{Schmid04, Shimizu21,dao24}. 
In contrast, variation after particle number projection (VAPNP) can be handled more easily to efficiently optimize pairing correlations characterizing the seed Bogoliubov states. 
Still, VAPNP calculations are typically one order of magnitude more expensive computationally than HFB calculations.

Given that \(1^+\) states are the focus of the present study, allowing the Bogoliubov states to break time-reversal symmetry is mandatory. 
Due to the additional computational cost associated with the breaking of this symmetry, which is often compounded by the inclusion of triaxiality\footnote{An appealing aspect, not explored here, would be to generate $1^+$ states without necessarily breaking axial symmetry given that angular momentum restoration of triaxially deformed states in large model spaces is numerically very involved.}, \(1^+\) states have not been addressed by state-of-the-art PGCM calculations. Here, two distinct sets of generator coordinates that constrain the seed Bogoliubov states to break time-reversal symmetry are considered to construct the PGCM states of interest (see details below).  The comparison between these two sets is meant to provide guidance for future calculations.

\begin{itemize}
    \item \textit{Set A} \\
The first set of generator coordinates is inspired by ingredients used in the linearized time-dependent mean-field equation,  and in particular in the finite amplitude method QRPA (FAM-QRPA). In FAM-QRPA, the operators driving the oscillations in the calculation of the photoabsorption strength function \(M1\) are nothing but the \(M^{10}\) and \(M^{11}\) components of the magnetic dipole operator.  It is thus natural to use those two components as collective coordinates to account for \(M1\) transitions in PGCM calculations. In order to deal efficiently with small transitions, the isovector component of the \(J_x\) operator $(J_x^{\mathrm{iv}})$ is added as a third generator coordinate.

The Bogoliubov states from Set A are obtained through a standard HFB minimization under constraint followed by a projection after variation (PAV) on all symmetry quantum numbers. The intervals of constraints are $\left[0,2\right]\mu_N$, $\left[0,3.5\right]\mu_N$ and $\left[0,5\right]$ for \(M^{10}\), \(M^{11}\) and  $J_x^{\mathrm{iv}}$, respectively. The step between two neighboring values is set to 0.5 for the three coordinates in order to ensure that the PGCM results remain practically unchanged by including additional intrinsic states.

PGCM calculations based on Set A have been carried out with the PAN@CEA generic solver~\cite{panacea}.

\item \textit{Set B} \\
The second set is based on the identification of the most relevant collective degrees of freedom of a nuclear Hamiltonian, namely multipole deformations and pairing~\cite{Dufour96}. These coordinates are usual choices in standard PGCM nuclear structure applications, and they do not target any particular observable apart from the energy. The deformations are not presently explored as generator coordinates but are simply left to optimize for each seed state. The VAPNP method is able to optimize like-particle 
(proton-proton and neutron-neutron) isovector pairing correlations such that, again, these parameters are not explored as generator coordinates. 

Eventually, the most relevant coordinates to be systematically explored are the proton-neutron (pn-) pairing amplitudes represented by $\delta^{T=0}_{pn}$ (isoscalar) and $\delta^{T=1}_{pn}$ (isovector) parameters~\cite{Bally21}. The former (latter) modulates the amount of pn-pairs coupled to $T=0$ and $J=1$ ($T=1$ and $J=0$) contained in  the Bogoliubov states that are expected to be particularly relevant to describe $N\approx Z$ nuclei such as $^{24}$Mg. The isoscalar component of the \(J_x\) operator $(J_x^{\mathrm{is}})$ is used  as the third generator coordinate. 

The Bogoliubov states from Set B are obtained through a VAPNP minimization under constraint followed by a projection after variation (PAV) on all symmetry quantum numbers.  The intervals of constraints are $\left[0,2.4\right]$ for  $\delta^{T=0,1}_{pn}$ and $\left[0,7\right]$ for $J_x^{\mathrm{is}}$, with a step
between two neighboring values of 0.4 for the pn-pairing degrees of freedom and 1 for the $J_x^{\mathrm{is}}$ coordinate.

PGCM calculations based on Set B have been carried out with the recently developed TAURUS suite~\cite{Bally21,Bally24}.
\end{itemize}

\section{Results}
\label{sec:results}

\subsection{Total energy surfaces}
\label{sec:results:tes}

\begin{figure}
\begin{center}
\includegraphics[width=\columnwidth]{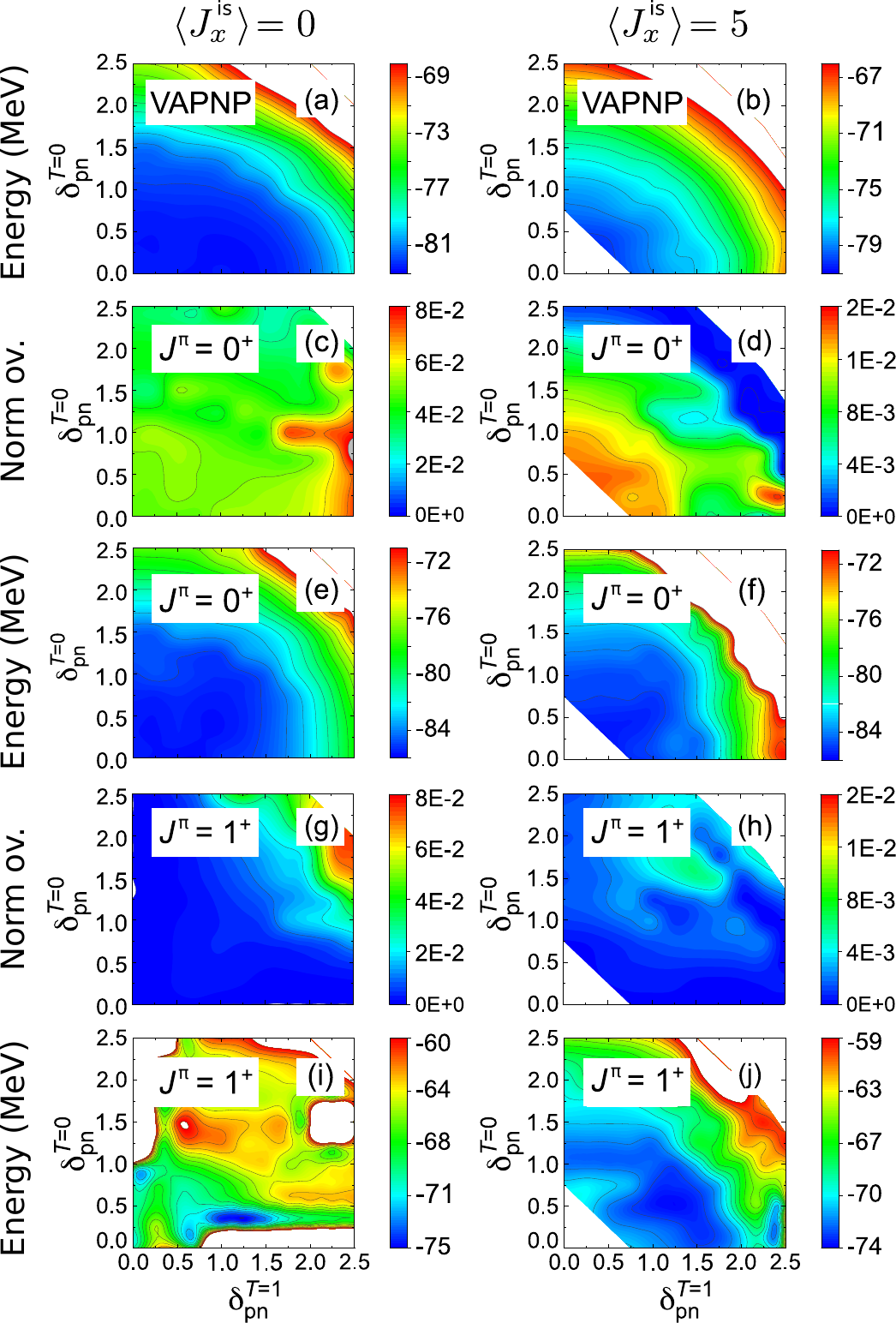}
\end{center}
\caption{Projected energy and norm surfaces in $^{24}$Mg as a function of the isovector and isoscalar pairing amplitudes $\delta^{T=1,0}_{pn}$ based on constrained HFB states associated with Set B. Fixed constraints $\langle J^{\mathrm{is}}_{x}\rangle = 0$ (left panels) and $\langle J^{\mathrm{is}}_{x}\rangle = 5$ (right panels) are employed. Panels: (a)-(b) VAPNP energy surfaces, (c)-(d) $J^\pi_\sigma=0^{+}_1$ projected norm overlap, (e)-(f) $0^{+}_1$ projected energy, (g)-(h) $1^{+}_1$ projected norm overlap and (i)-(j) $1^{+}_1$ projected energy.}
\label{fig:TES_setB}
\end{figure}

Figures~\ref{fig:TES_setB}(a)-(b) display the 
VAPNP energy surfaces as a function of the pn-pairing amplitudes for two illustrative values of the cranking constraint, 
i.e. $\langle J_{x}^{\mathrm{is}}\rangle$ = 0 and 5. The lowest energies are obtained for small pn-pairing amplitudes\footnote{The VAPNP equations do not converge for very small values of the pn-pairing amplitudes whenever $\langle J_{x}^{\mathrm{is}}\rangle\neq 0$ such that one cannot state with full confidence that the minimum is reached for null amplitudes in those cases. } such that the energy increases
steadily with $\delta^{T=0,1}_{pn}$ and is about 15\,MeV above the minimum when reaching the edges of the chosen intervals. 

The projected norm (energy) associated with each Bogoliubov state located on the total energy surface is displayed in Fig.~\ref{fig:TES_setB}(c)-(d) and (g)-(h) (Fig.~\ref{fig:TES_setB}(e)-(f) and (i)-(j)) for $J^\pi=0^{+}$ and $J^\pi=1^{+}$, respectively. 

First, projected norms different from zero indicate that the chosen generator coordinates are well suited to address states with the corresponding $J^\pi$ value. Results show that $J^\pi=0^{+}$ are of no concern given that they carry a large weight in all involved Bogoliubov states, except at large values of the pairing constraints for $\langle J_{x}^{\mathrm{is}}\rangle$ =  5. While it is more challenging for $1^{+}$ states, given that the corresponding projected norm overlaps are small over large areas of the 
$(\delta^{T=1}_{pn},\delta^{T=0}_{pn})$ plane, one obtains enough states with a projected norm different from zero to generate $1^{+}$ PGCM states in a controlled fashion (see below).  

Second, the span of projected energies provides a check of whether or not the range of values spanned by the constraints is appropriate to describe PGCM states of interest up to a chosen excitation energy. When further projected onto $J^\pi=0^{+}$ or $J^\pi=1^{+}$, the set of Bogoliubov states at play still covers a span of about 15\,MeV above the minimum of the projected total energy surface for both values of $\langle J_{x}^{\mathrm{is}}\rangle$. The topology of the $J^\pi=0^{+}$ projected energy surfaces is similar to their respective VAPNP counterpart, although the energy gained by the angular momentum restoration produces a deeper well. For $J^\pi=1^{+}$, the two projected total energy surfaces display a more discontinuous behavior, especially for $\langle J_{x}^{\mathrm{is}}\rangle=0$. 

Eventually, the set of constrained HFB states obtained based on Set B is appropriate to converge the $J^\pi=0^{+}$ PGCM ground state and $J^\pi=1^{+}$ PGCM states up to about 25\,MeV excitation energy discussed in the following. The same conclusion holds when employing HFB states from Set A.

\subsection{Resolution of the HWG equations}
\label{sec:results:hwg}

The next step corresponds to solving the HWG equation (Eq.~(\ref{eq:hwg})) separately for $J^\pi=0^{+}$ and $J^\pi=1^{+}$. The HWG equation, a generalized secular problem, is reduced to a standard diagonalization problem by canonically orthonormalizing the non-orthogonal and potentially linearly dependent projected Bogoliubov states
(see, e.g., Ref.~\cite{Martinez-Larraz22} and references therein for details). More specifically, the eigenvalues and eigenvectors of the projected norm overlap matrix (see Eq.~(\ref{eq:norm_kernels})) are used to extract a set of orthonormal states, the so-called natural basis of the collective space, onto which the PGCM states (Eq.~(\ref{eq:pgcm})) 
are expanded. At first, the natural basis may still contain strict and/or approximate 
linear dependencies (because of the machine numerical precision). Those are effectively removed by
\begin{enumerate}
\item ordering the natural basis states from the largest to the smallest eigenvalues of the norm overlap matrix,
\item diagonalizing successively the Hamiltonian matrices defined using an increasing number of natural basis states,
\item representing the corresponding PGCM eigenenergies as a function of such a number of natural basis states. 
\end{enumerate}
Proceeding in this way, linear dependencies manifest as abrupt drops in the lowest eigenvalue of the Hamiltonian matrix associated with a given $J^\pi$ of interest. Therefore, the converged HWG solutions are obtained from the Hamiltonian diagonalization performed just before the onset of such a numerical instability.

\begin{figure}
\begin{center}
\includegraphics[width=\columnwidth]{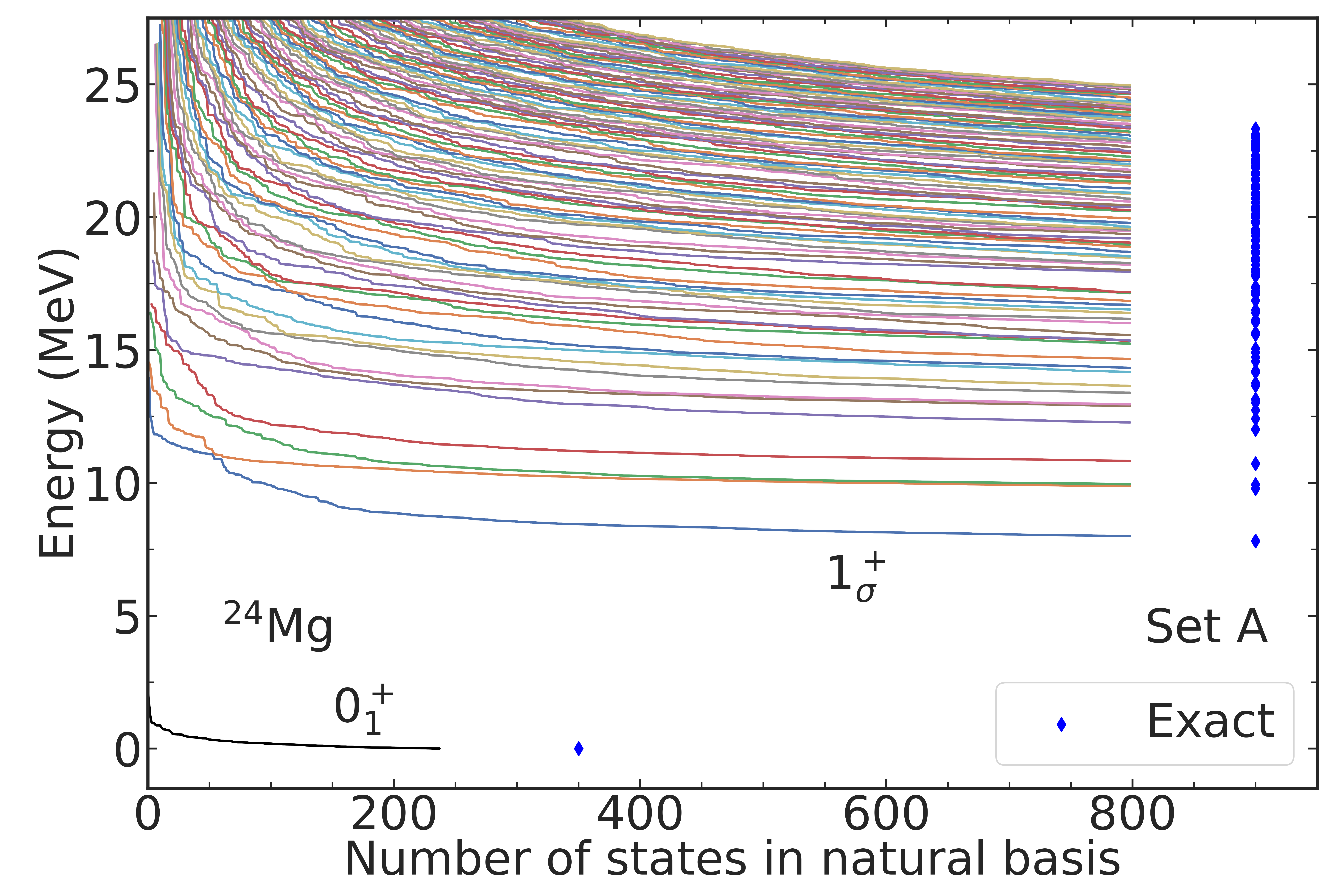}
\end{center}
\caption{PGCM energies of the $0^{+}_{1}$ ground state and $1_{\sigma}^{+}$ excited states in $^{24}$Mg as a function of the dimension of the natural basis, relative to the exact ground-state energy. PGCM energies from Set A are represented by full lines, exact shell-model ones are indicated by blue diamonds. 
}
\label{fig:pgcm_ener_nat_setA}
\end{figure}

\begin{figure}
\begin{center}
\includegraphics[width=\columnwidth]{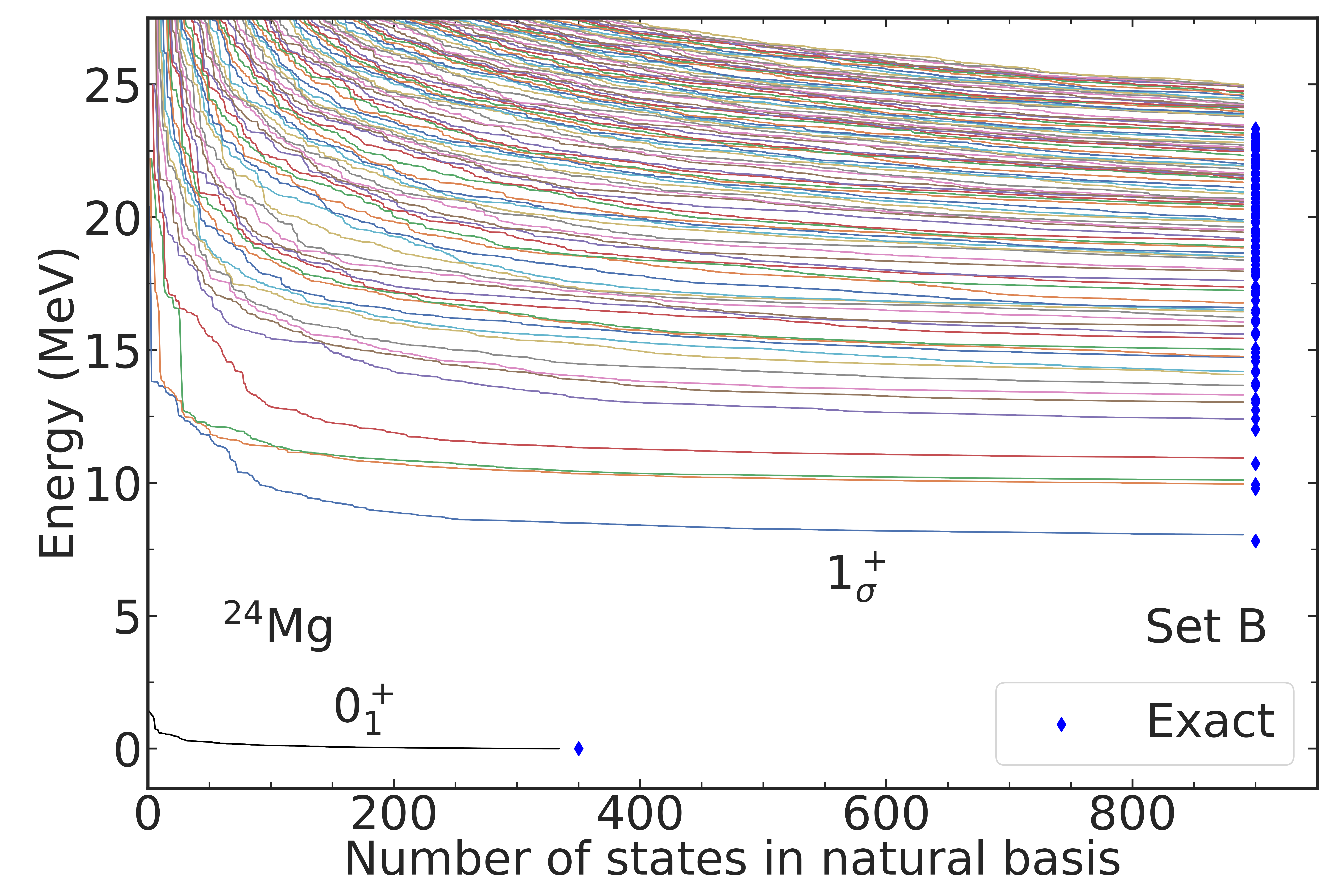}
\end{center}
\caption{{The same as in Fig.~\ref{fig:pgcm_ener_nat_setA} but for Set B. 
} 
\label{fig:pgcm_ener_nat_bas}}
\end{figure}

The PGCM $J^\pi_\sigma=0_{1}^{+}$ ground-state energy and all $J^\pi_\sigma=1_{\sigma}^{+}$ excited energies are displayed in Fig.~\ref{fig:pgcm_ener_nat_setA} (Fig.~\ref{fig:pgcm_ener_nat_bas}) for Set A (B) relative to the exact ground-state energy as a function of the number of states retained in the natural basis.
The $0_{1}^{+}$ and the lowest $1_{\sigma}^{+}$ energies quickly display a plateau as more natural basis states are added. Thus, the energy gain provided by natural states carrying increasingly smaller eigenvalues of the norm matrix becomes rapidly negligible. While reaching a plateau is more challenging for higher-lying $J^\pi_\sigma=1_{\sigma}^{+}$ states, it is possible to define convincingly converged energies in the present calculation for all $J^\pi_\sigma=1_{\sigma}^{+}$ before hitting a numerical instability. Eventually, the PGCM energies obtained with both Sets A and B compare very favorably with the exact ones also displayed in Figs.~\ref{fig:pgcm_ener_nat_setA} and \ref{fig:pgcm_ener_nat_bas}, especially for the lowest-energy states. 

The largest number of natural basis states that can be employed for Set A before linear dependencies appear is $240$ (out of $440$ projected Bogoliubov states) and $800$ (out of $3\times 440=1320$ projected Bogoliubov states\footnote{The factor of $3$ comes from the three allowed $K$ values for $J=1$.}) for $J^\pi=0^{+}$ and $J^\pi=1^{+}$, respectively.
With Set B, one can reach 320 states (out of 342 projected Bogoliubov states) and 890 states (out of 1026 projected Bogoliubov states) for $J^\pi=0^{+}$ and $J^\pi=1^{+}$, respectively. It thus appears that Set B is more efficient, i.e. linear dependencies among the set of projected Bogoliubov states are less important such that one can employ a large number of natural basis states starting from a smaller number of constrained Bogoliubov states\footnote{Given that the cost of PGCM calculations scales quadratically with the number of constrained Bogoliubov states, the capacity to reduce the number of constrained Bogoliubov states can become critical in large-scale applications.}. As a matter of fact, it allows the PGCM energies to be slightly better converged with Set B than with Set A.

Eventually, the number of natural-basis states employed in the present PGCM calculations is roughly 1/3 of the size of the complete Hilbert spaces used in the shell-model calculations (see Sec.~\ref{sec:results:obs}). 

\begin{figure*}
\begin{center}
\includegraphics[width=0.8\textwidth]{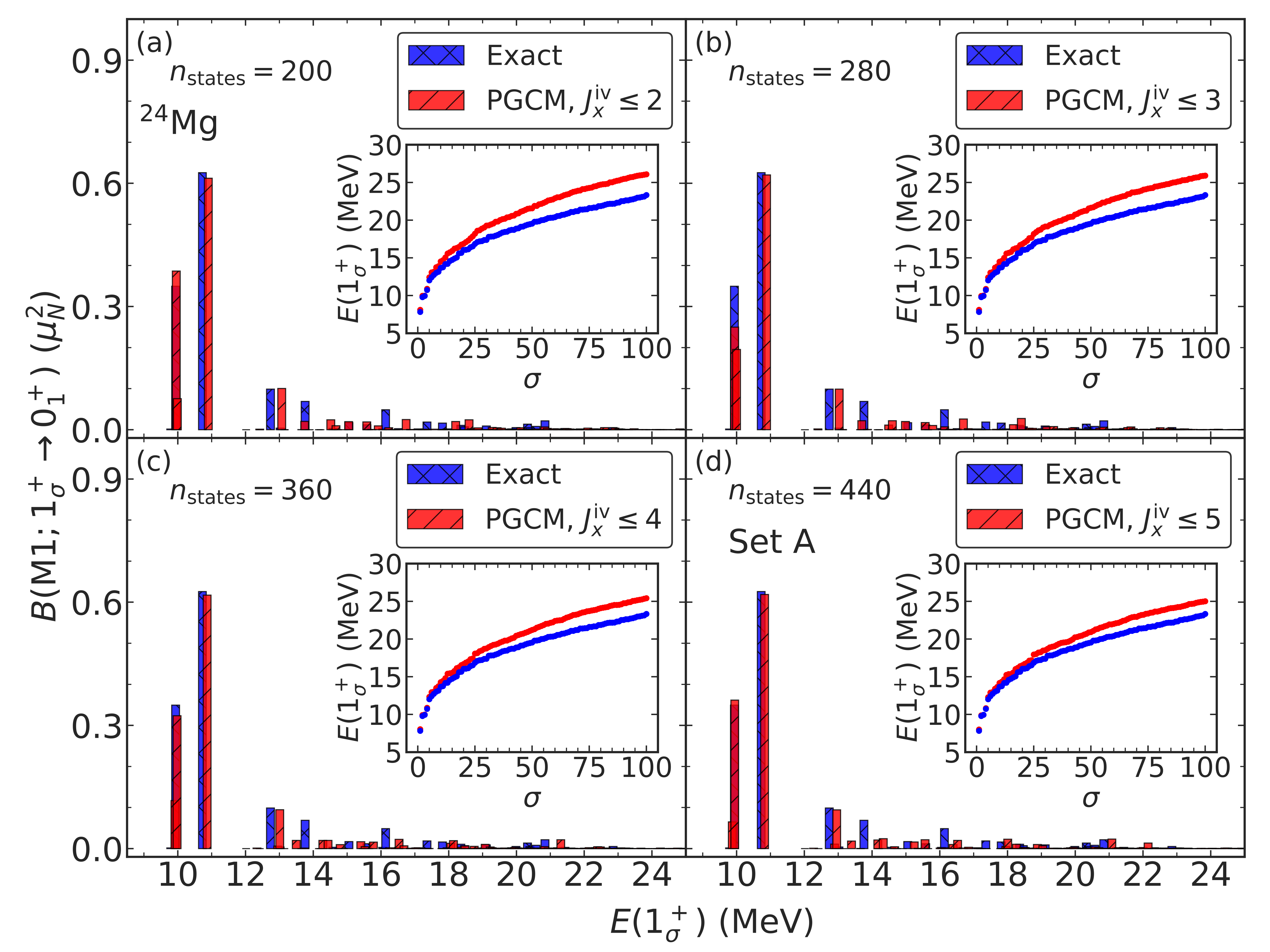}
\end{center}
\caption{Magnetic dipole transition probabilities, $B(M1; 1_{\sigma}^{+}\rightarrow 0^{+}_{1})$, as a function of the excitation energy $E(1^+_\sigma)$ in $^{24}$Mg.
Exact diagonalization (blue bars) and PGCM calculations with Set A (red bars) using different maximum values for the cranking constraint are shown. 
The comparison of the excitation energies of the lowest one hundred $1_{\sigma}^{+}$ states is shown in the insets.}
\label{fig:pgcm_bm1_setA}
\end{figure*}

\subsection{$E(1^{+})$ energies, $B(M1)$ transition strengths, and magnetic dipole moments
in $^{24}$Mg}
\label{sec:results:obs}

The present section focuses on the observables of interest in $^{24}$Mg. Placing 4 protons and 4 neutrons in the $sd$-shell orbitals one can construct 1161 and 3096
states coupled to $J^\pi=0^+$ and $J^\pi=1^+$, respectively.
The shell-model diagonalization, using the Lanczos algorithm, 
is presently carried out until the convergence of the ground state and the first 100 $1^+$ states reaching up to $\sim23$MeV excitation energy. Acting with the transition operator on an initial shell-model state and calculating its norm \cite{RMP} one obtains the sum rule of the operator in the model space (total strength). It could be checked that the strength carried by these 100 $1^+$ states $\sum_{i=1,100}B(M1; 1^+_i\rightarrow 0^+)=1.38\mu_N^2$ amounts to about 98\% of the total $B(M1)$ strength (1.41$\mu_N^2$).

\begin{figure*}
\begin{center}
\includegraphics[width=0.8\textwidth]{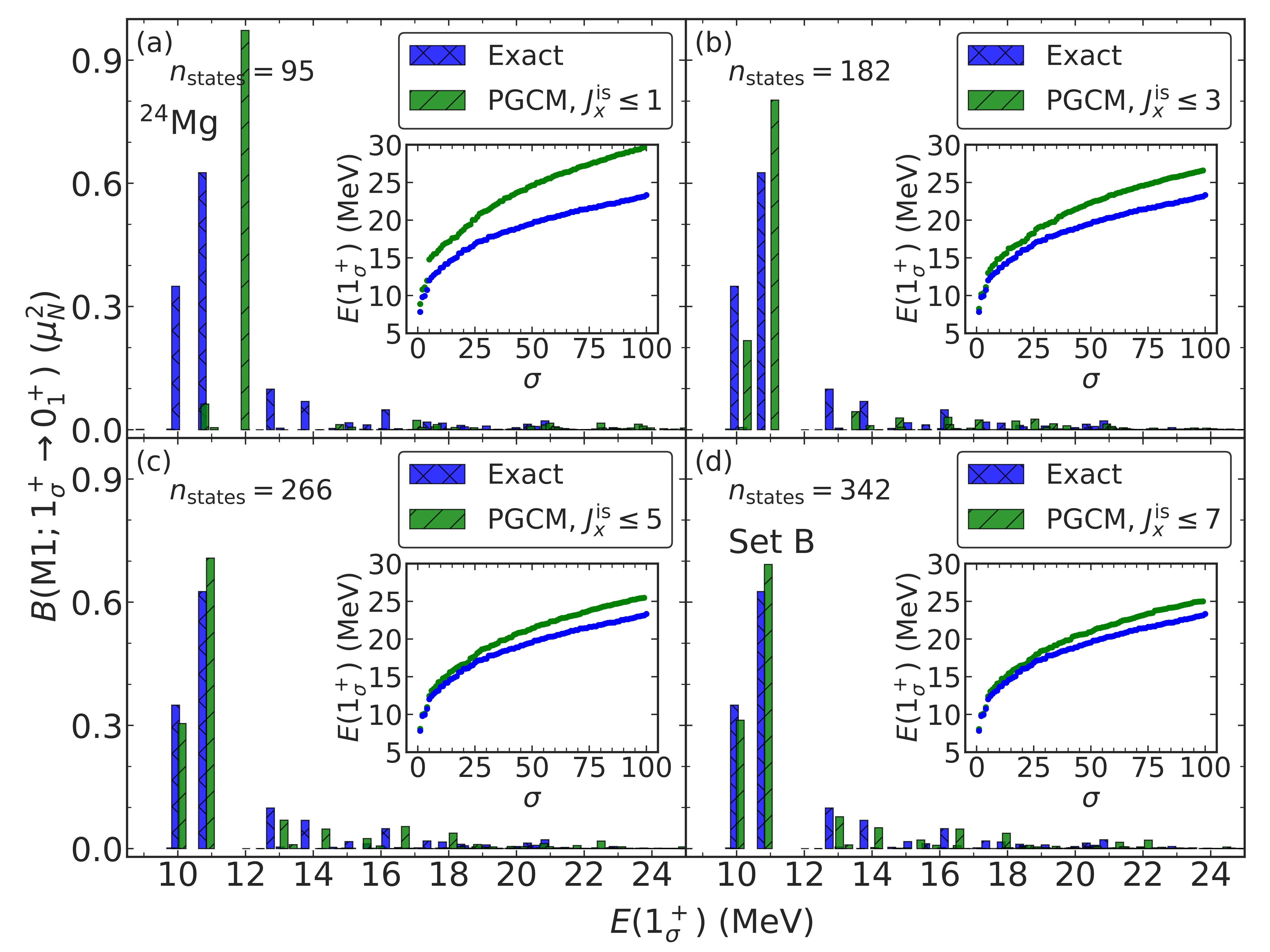}
\caption{The same as Fig.~\ref{fig:pgcm_bm1_setA} but for Set B (green bars).
}
\label{fig:pgcm_bm1_1plus}
\end{center}
\end{figure*}

Figure~\ref{fig:pgcm_bm1_setA} (\ref{fig:pgcm_bm1_1plus}) compares the PGCM $B(M1)$ strength function obtained for Set A (B) against the exact one. An inset also compares the excitation energies of the $1^{+}$ states themselves. 
Given that the number of constrained HFB states presently employed could be prohibitively large in full-space PGCM applications, the impact of limiting the initial set of constrained HFB states is also investigated. This is done by  using  $4$ increasing maximum values for the  constraint on $(J_x^{\mathrm{iv}})$ ($(J_x^{\mathrm{is}})$). The corresponding number of constrained HFB states is indicated in each panel. Table \ref{tab-rmsd} shows the corresponding root mean square deviation (RMSD) of $B(M1)$ transition values between PGCM and SM for each of the calculations presented in Figs. \ref{fig:pgcm_bm1_setA} and \ref{fig:pgcm_bm1_1plus}.

\begin{table}
\caption{Root mean square deviation (RMSD) of the magnetic dipole transition probabilities, $B(M1; 1_{\sigma}^{+}\rightarrow 0^{+}_{1})$, between exact diagonalization and PGCM calculations with Sets A and B. The deviation is displayed for sets with different maximum values for the cranking constraint indicated by the corresponding panel of Figs.~\ref{fig:pgcm_bm1_setA} and~\ref{fig:pgcm_bm1_1plus}.\label{tab-rmsd}}
\begin{tabular}{cT{1cm}T{1cm}T{1cm}T{1cm}}
\hline
    & \multicolumn{4}{c}{RMSD ($\mu_N^2$)} \\
panel & (a) & (b) & (c) & (d)\\
\hline
PGCM Set A & 0.051 & 0.034 & 0.015 & 0.011\\ 
\hline
PGCM Set B & 0.051 & 0.028 & 0.020 & 0.019\\ 
\hline
\end{tabular}
\end{table}

As a function of the excitation energy, exact results display two first $1^{+}$ states with small $B(M1)$ strength, followed by two states (at $\sim10.0$ and $~10.7$ MeV) with the largest transition probabilities. There are three more states with $B(M1)\geq 0.08$ at 12.7, 13.7 and 16.1 MeV before a quasi continuum of $1^{+}$ states appears (see insets). 
For both Sets A and B, the PGCM calculations display the same pattern, even though excitation energies of $1^{+}$ states are very well reproduced in the low-energy range and tend to be more and more overestimated towards higher energies. Still, both $M1$ transitions and  $1^{+}$ excitation energies improve significantly as the number of constrained Bogoliubov states is increased by adding more cranking configurations. As one can see in Tab. \ref{tab-rmsd}, however, including $J_x^{\mathrm{is}}$ values above 5 in Set B does not lead to substantial changes (see the difference between calculations (c) and (d)). Interestingly, if one is solely interested in the dominant contributions to the excitation strength, a decent reproduction is already obtained for both sets using about 200 constrained HFB states.

\begin{figure}
\includegraphics[width=1.0\columnwidth]{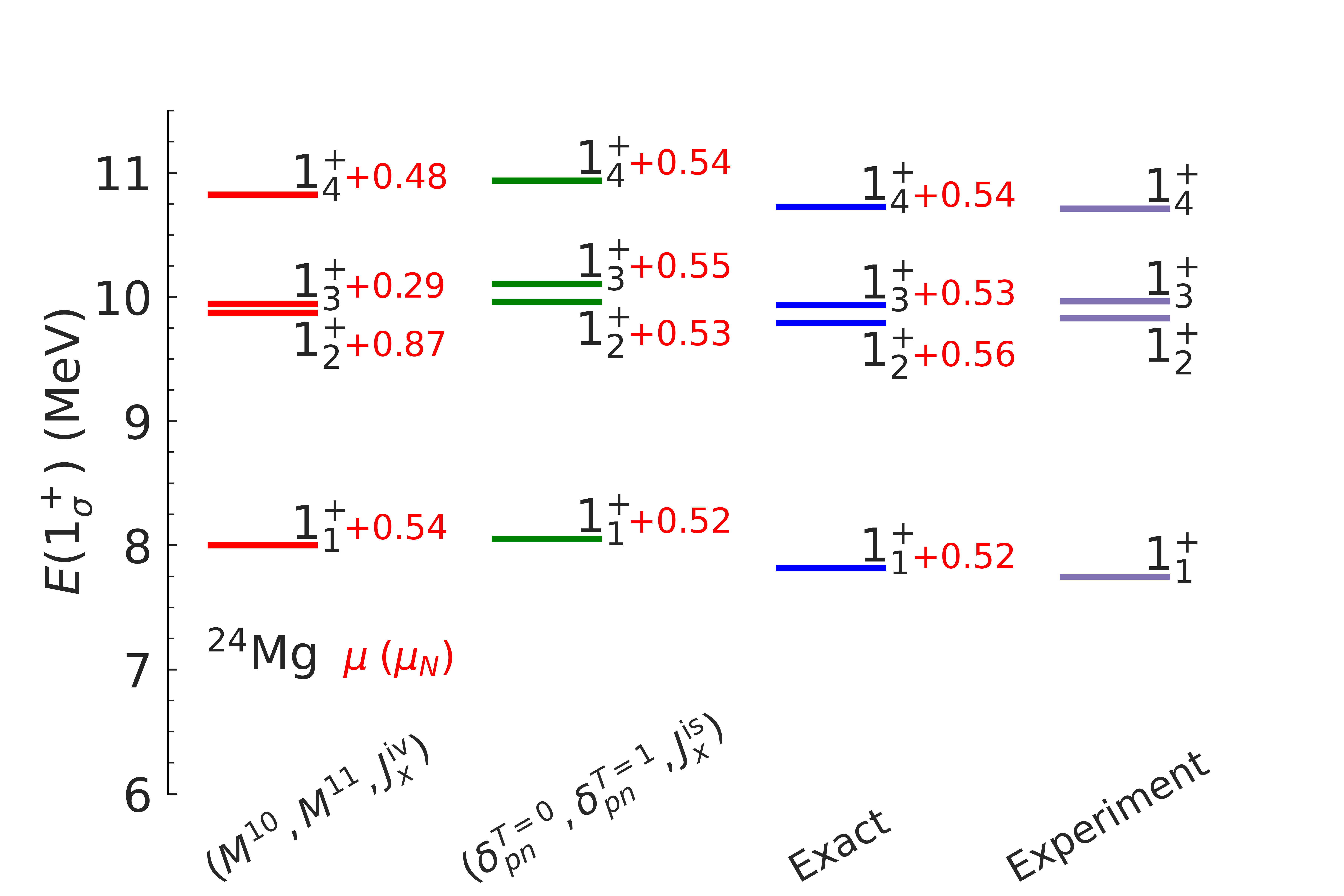}
\caption{Excitation energies and magnetic dipole moments of the four $1^+$ states of ${}^{24}$Mg obtained from PGCM calculations with Set A (red) and Set B (green), exact diagonalization (blue) and compared to available experimental data (purple) \cite{NNDC}. The excitation energies are given with respect to the corresponding ground-state energies of $-86.88$ MeV, $-87.08$ MeV and $-87.10$ MeV for PGCM Set A, Set B and exact diagonalization, respectively. 
}
\label{fig-spec}
\end{figure}

Figure \ref{fig-spec} compares the four lowest PGCM $1^{+}$ excitation energies and the associated magnetic dipole moment to the exact ones and the known experimental data\footnote{No experimental information on magnetic dipole moments of the $1^+$ states in $^{24}$Mg is available.}. First, PGCM $1^{+}$ excitation energies reproduce well the exact ones for both Sets A and B. While it is also the case for the magnetic moment of the first $1^{+}$ state, there is a large discrepancy in the prediction from Set A for the magnetic moments of the (too) closely-lying $1^+_{2,3}$ states. The agreement is again better for the $1^+_{4}$ that carries most of the $B(M1)$ strength. Set B, on the other hand, reproduces almost exactly the  magnetic moments of the four lowest  $1^{+}$ states. 

\begin{figure}
\begin{center}
\includegraphics[width=\columnwidth]{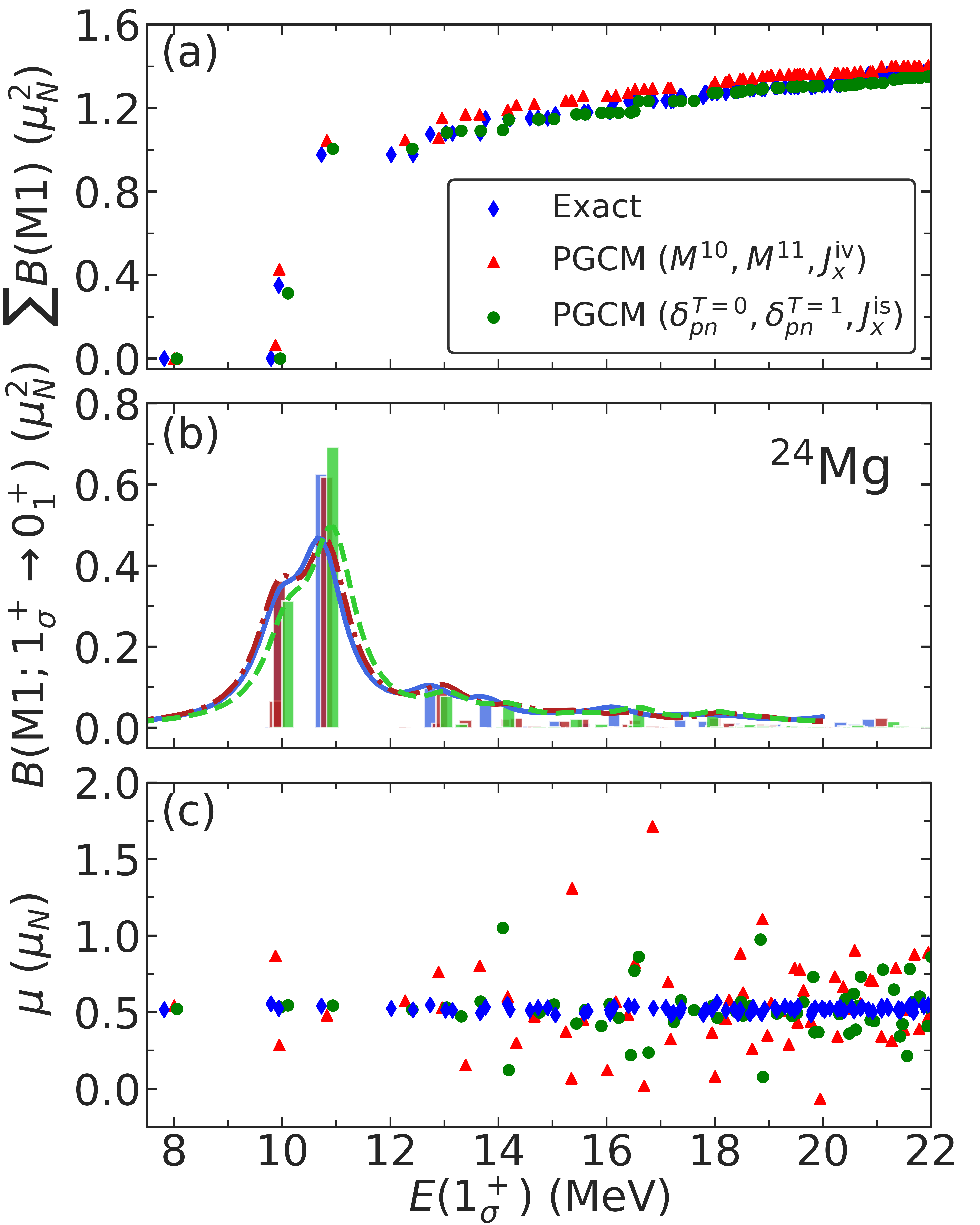}
\end{center}
\caption{(a) Cumulated $B(M1)$ strength, (b) discrete and convoluted $B(M1)$ strength functions, and (c) $1_{\sigma}^{+}$ magnetic 
dipole moments as a function of the excitation energy $E(1^+_\sigma)$ in $^{24}$Mg.
Results obtained from exact diagonalization (blue symbols/lines) are compared to PGCM calculations with Sets A (red symbols/lines) and B (green symbols/lines). 
}
\label{fig:pgcm_sum_lorentz_mu}
\end{figure}

The cumulated PGCM $B(M1)$ strength is shown in panel (a) of Fig.~\ref{fig:pgcm_sum_lorentz_mu} to reproduce very well the exact one: the total strength is 1.43 $\mu_N^2$ (1.39 $\mu_N^2$) for Set A (B), which is to be compared to 1.38 $\mu_N^2$ in the exact calculation. 

Similarly, the centroids and widths of the distributions shown in panel (b) of the same Figure are in excellent agreement; the PGCM centroid of 12.02\,MeV (12.39\,MeV) obtained for Set A (B) is close to the value of 12.10\,MeV of the exact calculation. The situation is similar for the widths equal to 3.06\,MeV, 3.26\,MeV and 3.08\,MeV, respectively. 

Magnetic dipole moments of all computed states are compared in panel (c) of Fig.~\ref{fig:pgcm_sum_lorentz_mu}. Exact values oscillate around 0.5$\mu_N$ without showing any particular trend with excitation energy. These almost constant values over the entire energy range are not well described within either of the PGCM calculations for $E(1^{+}_{\sigma})>13$ MeV, where the results scattered around $\mu\sim0.5\mu_{N}$. These states correspond to the high-level density region in the tail of the $B(M1)$ distribution, which is still well reproduced despite such inaccuracies in the structure of individual $1^+$ states. A closer look at the SM wave functions reveals a large mixing of various configurations, with the main components not exceeding $25\%$. The occupancies of the spherical orbits of the $1^+$ states carrying most of the strength in the exact diagonalization are shown in Table~\ref{tab-occup}
and they are compared to the occupations resulting from the PGCM calculation with $J_x^{\mathrm{is}}\leq7$ from Set B. As shown, the occupation of spherical orbits of the $0^+$ and $1^+$ states is nearly identical in the PGCM and SM calculations; 
still, small differences are present (see the comparison between the $1^+_{6}$ PGCM state and the SM $1^+_{7}$). This means that the variational wave functions do not necessarily capture the full complexity of the exact wave functions.


\begin{table}
\caption{Excitation energies, magnetic dipole moments and mean occupation of spherical orbitals obtained from exact diagonalization and PGCM calculations (Set B with cranking values up to $J_x^{\mathrm{is}}\leq7$). The reported states include the ground state and the $1^+$ states carrying most of the $B(M1)$ strength in $^{24}$Mg. \label{tab-occup}}
\begin{tabular}{ccccccc}
\hline
&$J^\pi$ & $E_{\rm exc}(\rm MeV)$ & $\mu (\mu_N)$& $0d_{5/2}$ & $1s_{1/2}$ & $0d_{3/2}$\\
\hline
Exact &$0^+$ & 0.0 & -&3.04 & 0.43 & 0.53 \\
&$1^+_3$& 9.9& 0.52 &3.27 & 0.42 & 0.31 \\  
&$1^+_4$& 10.7 &0.54&2.88 & 0.68 & 0.44 \\
&$1^+_7$& 12.7 &0.55&2.86 & 0.68 & 0.46 \\
&$1^+_{11}$& 13.8&0.53&3.03 & 0.53 & 0.44\\ 
\hline
PGCM  &$0^+$ &0.0 &-   &3.05 & 0.43 & 0.52 \\
&$1^+_3$& 10.1&0.55    &3.27 & 0.42 & 0.31 \\  
&$1^+_4$& 10.9 &0.54   &2.90 & 0.64 & 0.46 \\
&$1^+_6$& 13.0 &0.53   &2.85 & 0.58 & 0.57 \\
&$1^+_{10}$& 14.2 &0.60 &3.05 & 0.52 & 0.43\\ 
\hline
\end{tabular}
\end{table}

\begin{figure}
\includegraphics[width=\columnwidth]{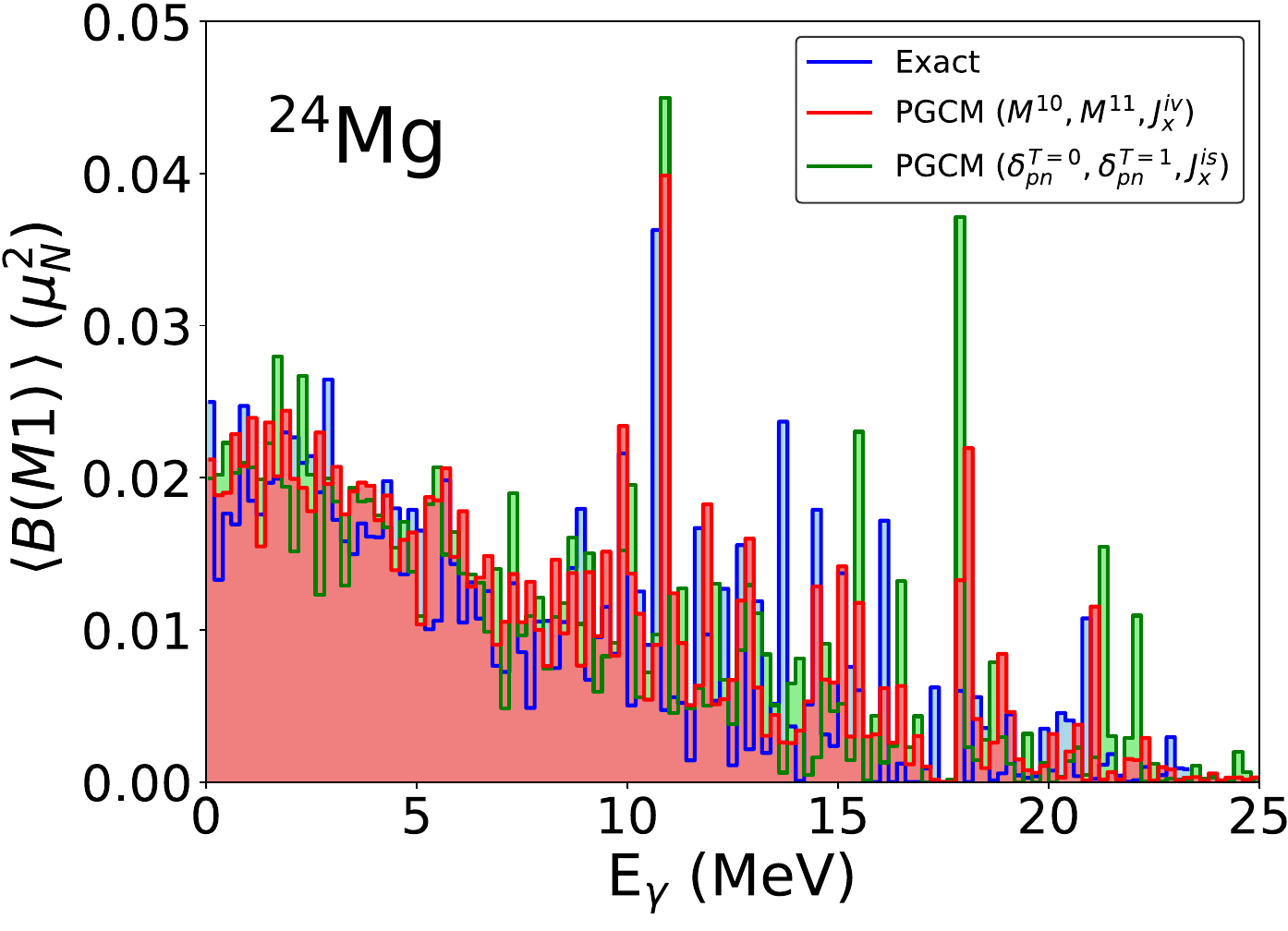}
\caption{Averaged $B(M1)$ de-excitation strength in $^{24}$Mg. $1^+_{\sigma} \rightarrow 1^+_{\sigma'} $ and $1^+_{\sigma} \rightarrow 0^+_{1} $ transitions for the first 100 $1^+$ states are included in the calculations, using a binning of 0.2 MeV for the transition energy $E_\gamma$.
Exact results (blue) are compared to PGCM calculations with Sets A (red) and B (green). }
\label{fig-LEE}
\end{figure}

Finally, the de-excitation $B(M1)$ strength is presented in Fig.~\ref{fig-LEE}. Following the approach used in previous shell-model studies~\cite{Schwengner-Mo, Brown-fe56, Sieja-PRL, Sieja2018}, 
which successfully reproduce the low-energy enhancement effect, 
the averaged $\langle B(M1)\rangle $ strength per $\gamma$-energy bin of 0.2~MeV width is computed. In these calculations, all $M1$ transitions are included, i.e. transitions connecting $1^+$ states to the ground state, as well as those occurring between $1^+$ excited states. This amounts to $5050$ transition matrix elements in total. The ability to compute the transitions between highly excited states is crucial
to capture the structural effects emerging at the lowest $\gamma$ energies in de-excitation strength functions, such as the LEE. While earlier shell-model calculations averaged over many spin values to get the de-excitation strength, it has been demonstrated that the latter is not significantly influenced by the range of spins considered. Instead, the LEE effect is predominantly governed by the level density. 

As shown in Fig.~\ref{fig-LEE}, both PGCM calculations exhibit an excellent agreement with the SM reference result, in particular regarding the dependence of $\langle B(M1)\rangle$ on $E_\gamma$. The prominent peak around $E_\gamma$=10~MeV corresponds to large spin-flip transitions to the $0^+$ ground state, which remain visible in the averaged $M1$ strength. Despite the relatively limited number of states included in the present calculations, a distinct upward trend of $\langle B(M1)\rangle$ toward $E_\gamma=0$ is clearly observed. The magnitude of the upbend is accurately captured by both PGCM calculations, despite differences in the predicted structure of the $1^+$ states above 13~MeV as discussed above. This outcome is consistent with the observation that the LEE effect is primarily driven by the level density, which is similar in both PGCM and shell-model calculations. As anticipated in Ref.~\cite{Frosini2023}, the PGCM is shown to be a suitable framework to accurately model the structural effects that influence the $\gamma$-ray strength functions, particularly in the low-energy regime.

\subsection{$B(M1)$ strengths in other $sd$-shell nuclei}
\label{sec:results:other}

As a final part of this study, PGCM calculations of $M1$ strength functions are extended to other \textit{sd}-shell nuclei using Set B. Thus, 350 constrained Bogoliubov states corresponding to $J_x^{\mathrm{is}}\leq5$ are used to study two $N=Z$ nuclei ($^{20}$Ne, $^{28}$Si) and two magnesium isotopes ($^{26}$Mg,$^{28}$Mg) with $N=Z+2$ and $N=Z+4$.

\begin{figure*}
\begin{center}
\includegraphics[width=0.85\textwidth]{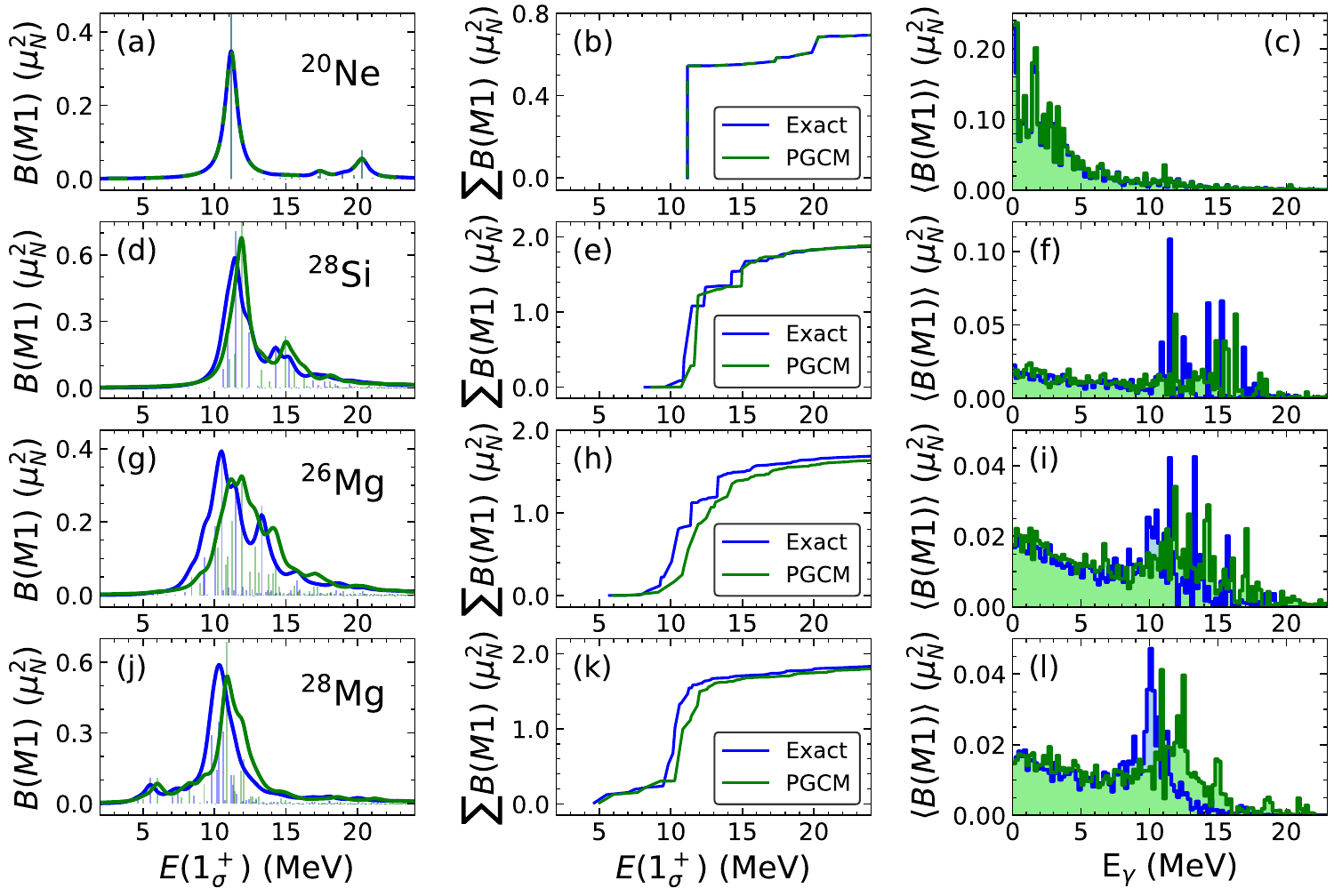}
\end{center}
\caption{Convoluted, cumulated and averaged $B(M1)$ strength functions for: $^{20}$Ne (a)-(c), $^{28}$Si (d)-(f), $^{26}$Mg (g)-(i) and $^{28}$Mg (j)-(l). PGCM results based on Set B (green) are compared to shell-model results (blue). }
\label{fig:other_nuclei}
\end{figure*}

While Table \ref{tab-othernuc} lists ground-state energies and properties of the ground-state $M1$ distributions, the corresponding strength functions are shown in panels (a), (d), (g), (j) of Fig.~\ref{fig:other_nuclei}. Panels (b), (e), (h), (k) display the cumulative $B(M1)$ strength, while panels (c), (f), (i), (l) show the averaged $\langle B(M1)\rangle$ transition values, calculated in the same manner as for $^{24}$Mg (see Fig.~\ref{fig-LEE}). 

With 2 protons and 2 neutrons above a core of $^{16}$O, the PGCM and shell-model results for $^{20}$Ne are found to be equivalent, indicating that the small valence space considered spans a less complex space of $1^+$ states compared to $^{24}$Mg. Staying on the $N=Z$ line with $^{28}$Si, there are now 6 protons and 6 neutrons distributed within the \textit{sd} valence space. The general behavior of the $M1$ distribution is reproduced, even though  $1^+$ states' excitation energies are slightly overestimated, resulting in a centroid deviation of 0.25 MeV, see Table \ref{tab-othernuc}, and a lack of fragmentation in the peak structure around 15 MeV. Still, the PGCM calculation displays an excellent agreement with the exact diagonalization regarding the total accumulated strength for the transitions from the first 100 $1^+$ states to the ground state.

For isotopes beyond $^{24}$Mg, PGCM results are found to reproduce rather well the exact ones for both the general shape of the $B(M1)$ distribution and the cumulated $M1$ strength even though some discrepancy remains, especially for $^{26}$Mg. This is not too surprising given that the generator coordinates defining Set B were selected with a focus on $N=Z$ nuclei. In fact, $^{26}$Mg exhibits a total energy surface that is rather soft against triaxial deformations, a feature that is not explored in the present calculations but could deserve more attention. Indeed, PGCM calculations for this nucleus display the least favorable results regarding the fragmentation of the strength and the total $M1$ strength of the five studied nuclei (see Fig.~\ref{fig:other_nuclei}(g-h) and Table \ref{tab-othernuc}). However, the disagreement for the latter is only 0.09 $\mu^2_N$, i.e. 5\%, away from the exact value.

Regarding averaged $M1$ strengths presented in panels (c), (f), (i) and (l) of Fig.~\ref{fig:other_nuclei}, the agreement is particularly good for transition energies up to 8-10 MeV, which is mostly dominated by the transitions between $1^+$ states in the quasi-continuum.
Thus, the fact that the PGCM description of the LEE is similar to the one provided by exact result for the five $sd$ nuclei under consideration is a key result of the present work. Still, similarly to what was shown in the insets of Fig.~\ref{fig:pgcm_bm1_1plus} for $^{24}$Mg, the higher-energy states are less accurately reproduced than the lower-energy ones for $^{28}$Si, $^{26}$Mg and $^{28}$Mg. This implies greater discrepancies for transitions with large $E_{\gamma}$ that connect states from the high- to low-energy regimes.

\begin{table}
\caption{Ground state binding energy ($E(0^+_1)$), total strength ($S_0$), centroid ($\bar S$) and width ($\Delta S$) of the $M1$ strength distribution  presented in panels (a),(d),(g) and (j) of Fig.~\ref{fig:other_nuclei} (for $^{20}$Ne, $^{28}$Si, $^{26}$Mg, $^{28}$Mg nuclei), obtained in shell-model (exact) and in PGCM calculations.
\label{tab-othernuc}}
\begin{tabular}{ccccccc}
\hline
& Calculation & $E(0^+_1)(\rm MeV)$  & $S_0(\rm MeV)$ & $\bar{S}(\rm MeV)$ & $\Delta S (\rm MeV)$\\
\hline
$^{20}$Ne & Exact & -40.47 & 0.70 & 13.07 & 3.76 \\
          & PGCM  & -40.47 & 0.70 &13.08 & 3.77  \\  
\hline
$^{28}$Si & Exact & -135.86 & 1.90& 13.06 & 3.10 \\
          & PGCM  & -135.64 & 1.89& 13.29 & 2.55  \\  
\hline

$^{26}$Mg & Exact & -105.52 & 1.72 & 12.14 & 3.35\\
          & PGCM  & -105.34 & 1.63 & 12.84 & 2.54 \\  
\hline
$^{28}$Mg & Exact & -120.49 & 1.85 & 10.87 & 3.34 \\
          & PGCM  & -120.38 & 1.78 & 11.27 & 2.71 \\  
\hline
\end{tabular}
\end{table}

\section{Conclusions}
\label{sec:conclusion}

The projected generator coordinate method has essentially never been used to systematically address spectroscopic properties of $1^+$ states due to the necessity
to break time-reversal symmetry when generating the adiabatic mean-field states involved in the configuration mixing. The present study demonstrated, for a selection of $sd$-shell nuclei, that the PGCM is a reliable tool to reproduce valence space shell-model results without requiring the full diagonalization of the valence-space Hamiltonian.

Two sets of generator coordinates were explored, each guided by a different physical intuition, to investigate in detail $M1$ properties in $^{24}$Mg.  
Both sets successfully reproduced exact results across observables of interest, including excitation energies and magnetic dipole moments of the lowest-lying $1^+$ states, level density, $B(M1)$ strength distributions and for the first time, the LEE of the $M1$ de-excitation strength.
This suggests that the specific choice of generator coordinates is less critical, provided that the selected degrees of freedom are physically meaningful
and the underlying HFB states effectively break time-reversal symmetry.
It was further demonstrated that the same prescription can 
generate $M1$ strengths in good accord with exact results for both $N=Z$ nuclei and nuclei with neutron-proton asymmetry.

This study establishes a foundation for more systematic testing of the $M1$ strengths, either within valence spaces using shell-model Hamiltonians, or in large configuration spaces based on energy density functionals or realistic nuclear Hamiltonians obtained from chiral effective field theory.

\begin{acknowledgements}
KS acknowledges funding from the Interdisciplinary Thematic Institute QMat, as part of the ITI 2021-2028 program of the University of Strasbourg, CNRS and Inserm, supported by the IdEx Unistra (ANR 10-IDEX-0002), and by the SFRI STRAT’US (ANR 20-SFRI-0012) and EUR QMAT (ANR 17-EURE-0024) projects under the framework of the French Investments for the Future Program. JML and TRR acknowledge funding from the Spanish MICIN under PRE2019-
088036 and PID2021-127890NB-I00, and support from GSI-Darmstadt computing facilities.

\end{acknowledgements}

\bibliography{bibliography.bib}

\begin{thebibliography}{49}%
\makeatletter
\providecommand \@ifxundefined [1]{%
 \@ifx{#1\undefined}
}%
\providecommand \@ifnum [1]{%
 \ifnum #1\expandafter \@firstoftwo
 \else \expandafter \@secondoftwo
 \fi
}%
\providecommand \@ifx [1]{%
 \ifx #1\expandafter \@firstoftwo
 \else \expandafter \@secondoftwo
 \fi
}%
\providecommand \natexlab [1]{#1}%
\providecommand \enquote  [1]{``#1''}%
\providecommand \bibnamefont  [1]{#1}%
\providecommand \bibfnamefont [1]{#1}%
\providecommand \citenamefont [1]{#1}%
\providecommand \href@noop [0]{\@secondoftwo}%
\providecommand \href [0]{\begingroup \@sanitize@url \@href}%
\providecommand \@href[1]{\@@startlink{#1}\@@href}%
\providecommand \@@href[1]{\endgroup#1\@@endlink}%
\providecommand \@sanitize@url [0]{\catcode `\\12\catcode `\$12\catcode `\&12\catcode `\#12\catcode `\^12\catcode `\_12\catcode `\%12\relax}%
\providecommand \@@startlink[1]{}%
\providecommand \@@endlink[0]{}%
\providecommand \url  [0]{\begingroup\@sanitize@url \@url }%
\providecommand \@url [1]{\endgroup\@href {#1}{\urlprefix }}%
\providecommand \urlprefix  [0]{URL }%
\providecommand \Eprint [0]{\href }%
\providecommand \doibase [0]{http://dx.doi.org/}%
\providecommand \selectlanguage [0]{\@gobble}%
\providecommand \bibinfo  [0]{\@secondoftwo}%
\providecommand \bibfield  [0]{\@secondoftwo}%
\providecommand \translation [1]{[#1]}%
\providecommand \BibitemOpen [0]{}%
\providecommand \bibitemStop [0]{}%
\providecommand \bibitemNoStop [0]{.\EOS\space}%
\providecommand \EOS [0]{\spacefactor3000\relax}%
\providecommand \BibitemShut  [1]{\csname bibitem#1\endcsname}%
\let\auto@bib@innerbib\@empty
\bibitem [{\citenamefont {Kopecky}()}]{SLO}%
  \BibitemOpen
  \bibfield  {author} {\bibinfo {author} {\bibfnamefont {J.}~\bibnamefont {Kopecky}},\ }\href@noop {} {}\bibinfo {howpublished} {in Handbook for calculations of nuclear reaction data. Reference Input Parameter Library (RIPL), IAEA-TEDOC-1034, 1998, Ch.6}\BibitemShut {NoStop}%
\bibitem [{\citenamefont {Capote}\ \emph {et~al.}(2009)\citenamefont {Capote}, \citenamefont {Herman}, \citenamefont {Obložinský}, \citenamefont {Young}, \citenamefont {Goriely}, \citenamefont {Belgya}, \citenamefont {Ignatyuk}, \citenamefont {Koning}, \citenamefont {Hilaire}, \citenamefont {Plujko}, \citenamefont {Avrigeanu}, \citenamefont {Bersillon}, \citenamefont {Chadwick}, \citenamefont {Fukahori}, \citenamefont {Ge}, \citenamefont {Han}, \citenamefont {Kailas}, \citenamefont {Kopecky}, \citenamefont {Maslov}, \citenamefont {Reffo}, \citenamefont {Sin}, \citenamefont {Soukhovitskii},\ and\ \citenamefont {Talou}}]{CAPOTE20093107}%
  \BibitemOpen
  \bibfield  {author} {\bibinfo {author} {\bibfnamefont {R.}~\bibnamefont {Capote}}, \bibinfo {author} {\bibfnamefont {M.}~\bibnamefont {Herman}}, \bibinfo {author} {\bibfnamefont {P.}~\bibnamefont {Obložinský}}, \bibinfo {author} {\bibfnamefont {P.}~\bibnamefont {Young}}, \bibinfo {author} {\bibfnamefont {S.}~\bibnamefont {Goriely}}, \bibinfo {author} {\bibfnamefont {T.}~\bibnamefont {Belgya}}, \bibinfo {author} {\bibfnamefont {A.}~\bibnamefont {Ignatyuk}}, \bibinfo {author} {\bibfnamefont {A.}~\bibnamefont {Koning}}, \bibinfo {author} {\bibfnamefont {S.}~\bibnamefont {Hilaire}}, \bibinfo {author} {\bibfnamefont {V.}~\bibnamefont {Plujko}}, \bibinfo {author} {\bibfnamefont {M.}~\bibnamefont {Avrigeanu}}, \bibinfo {author} {\bibfnamefont {O.}~\bibnamefont {Bersillon}}, \bibinfo {author} {\bibfnamefont {M.}~\bibnamefont {Chadwick}}, \bibinfo {author} {\bibfnamefont {T.}~\bibnamefont {Fukahori}}, \bibinfo {author} {\bibfnamefont {Z.}~\bibnamefont {Ge}}, \bibinfo {author} {\bibfnamefont {Y.}~\bibnamefont
  {Han}}, \bibinfo {author} {\bibfnamefont {S.}~\bibnamefont {Kailas}}, \bibinfo {author} {\bibfnamefont {J.}~\bibnamefont {Kopecky}}, \bibinfo {author} {\bibfnamefont {V.}~\bibnamefont {Maslov}}, \bibinfo {author} {\bibfnamefont {G.}~\bibnamefont {Reffo}}, \bibinfo {author} {\bibfnamefont {M.}~\bibnamefont {Sin}}, \bibinfo {author} {\bibfnamefont {E.}~\bibnamefont {Soukhovitskii}}, \ and\ \bibinfo {author} {\bibfnamefont {P.}~\bibnamefont {Talou}},\ }\href@noop {} {\bibfield  {journal} {\bibinfo  {journal} {Nuclear Data Sheets}\ }\textbf {\bibinfo {volume} {110}},\ \bibinfo {pages} {3107} (\bibinfo {year} {2009})},\ \bibinfo {note} {special Issue on Nuclear Reaction Data}\BibitemShut {NoStop}%
\bibitem [{RIP()}]{RIPL-3}%
  \BibitemOpen
  \href@noop {} {}\bibinfo {howpublished} {https://www-nds.iaea.org/RIPL-3/}\BibitemShut {NoStop}%
\bibitem [{\citenamefont {Larsen}\ \emph {et~al.}(2013)\citenamefont {Larsen} \emph {et~al.}}]{Larsen-Fe56}%
  \BibitemOpen
  \bibfield  {author} {\bibinfo {author} {\bibfnamefont {A.~C.}\ \bibnamefont {Larsen}} \emph {et~al.},\ }\href {\doibase 10.1103/PhysRevLett.111.242504} {\bibfield  {journal} {\bibinfo  {journal} {Phys. Rev. Lett.}\ }\textbf {\bibinfo {volume} {111}},\ \bibinfo {pages} {242504} (\bibinfo {year} {2013})}\BibitemShut {NoStop}%
\bibitem [{\citenamefont {Utsunomiya}\ \emph {et~al.}(2013)\citenamefont {Utsunomiya}, \citenamefont {Goriely}, \citenamefont {Kondo}, \citenamefont {Iwamoto}, \citenamefont {Akimune}, \citenamefont {Yamagata}, \citenamefont {Toyokawa}, \citenamefont {Harada}, \citenamefont {Kitatani}, \citenamefont {Lui}, \citenamefont {Larsen}, \citenamefont {Guttormsen}, \citenamefont {Koehler}, \citenamefont {Hilaire}, \citenamefont {P\'eru}, \citenamefont {Martini},\ and\ \citenamefont {Koning}}]{enh-Mo}%
  \BibitemOpen
  \bibfield  {author} {\bibinfo {author} {\bibfnamefont {H.}~\bibnamefont {Utsunomiya}}, \bibinfo {author} {\bibfnamefont {S.}~\bibnamefont {Goriely}}, \bibinfo {author} {\bibfnamefont {T.}~\bibnamefont {Kondo}}, \bibinfo {author} {\bibfnamefont {C.}~\bibnamefont {Iwamoto}}, \bibinfo {author} {\bibfnamefont {H.}~\bibnamefont {Akimune}}, \bibinfo {author} {\bibfnamefont {T.}~\bibnamefont {Yamagata}}, \bibinfo {author} {\bibfnamefont {H.}~\bibnamefont {Toyokawa}}, \bibinfo {author} {\bibfnamefont {H.}~\bibnamefont {Harada}}, \bibinfo {author} {\bibfnamefont {F.}~\bibnamefont {Kitatani}}, \bibinfo {author} {\bibfnamefont {Y.-W.}\ \bibnamefont {Lui}}, \bibinfo {author} {\bibfnamefont {A.~C.}\ \bibnamefont {Larsen}}, \bibinfo {author} {\bibfnamefont {M.}~\bibnamefont {Guttormsen}}, \bibinfo {author} {\bibfnamefont {P.~E.}\ \bibnamefont {Koehler}}, \bibinfo {author} {\bibfnamefont {S.}~\bibnamefont {Hilaire}}, \bibinfo {author} {\bibfnamefont {S.}~\bibnamefont {P\'eru}}, \bibinfo {author} {\bibfnamefont
  {M.}~\bibnamefont {Martini}}, \ and\ \bibinfo {author} {\bibfnamefont {A.~J.}\ \bibnamefont {Koning}},\ }\href {\doibase 10.1103/PhysRevC.88.015805} {\bibfield  {journal} {\bibinfo  {journal} {Phys. Rev. C}\ }\textbf {\bibinfo {volume} {88}},\ \bibinfo {pages} {015805} (\bibinfo {year} {2013})}\BibitemShut {NoStop}%
\bibitem [{\citenamefont {B\"urger}\ \emph {et~al.}(2012)\citenamefont {B\"urger} \emph {et~al.}}]{end-Sc43}%
  \BibitemOpen
  \bibfield  {author} {\bibinfo {author} {\bibfnamefont {A.}~\bibnamefont {B\"urger}} \emph {et~al.},\ }\href {\doibase 10.1103/PhysRevC.85.064328} {\bibfield  {journal} {\bibinfo  {journal} {Phys. Rev. C}\ }\textbf {\bibinfo {volume} {85}},\ \bibinfo {pages} {064328} (\bibinfo {year} {2012})}\BibitemShut {NoStop}%
\bibitem [{\citenamefont {Larsen}\ \emph {et~al.}(2012)\citenamefont {Larsen} \emph {et~al.}}]{Larsen-Ti44}%
  \BibitemOpen
  \bibfield  {author} {\bibinfo {author} {\bibfnamefont {A.~C.}\ \bibnamefont {Larsen}} \emph {et~al.},\ }\href {\doibase 10.1103/PhysRevC.85.014320} {\bibfield  {journal} {\bibinfo  {journal} {Phys. Rev. C}\ }\textbf {\bibinfo {volume} {85}},\ \bibinfo {pages} {014320} (\bibinfo {year} {2012})}\BibitemShut {NoStop}%
\bibitem [{\citenamefont {Larsen}\ and\ \citenamefont {Goriely}(2010)}]{Larsen2010}%
  \BibitemOpen
  \bibfield  {author} {\bibinfo {author} {\bibfnamefont {A.~C.}\ \bibnamefont {Larsen}}\ and\ \bibinfo {author} {\bibfnamefont {S.}~\bibnamefont {Goriely}},\ }\href {\doibase 10.1103/PhysRevC.82.014318} {\bibfield  {journal} {\bibinfo  {journal} {Phys. Rev. C}\ }\textbf {\bibinfo {volume} {82}},\ \bibinfo {pages} {014318} (\bibinfo {year} {2010})}\BibitemShut {NoStop}%
\bibitem [{osl()}]{oslo-website}%
  \BibitemOpen
  \href@noop {} {}\bibinfo {howpublished} {http://www.mn.uio.no/fysikk/}\BibitemShut {NoStop}%
\bibitem [{\citenamefont {Ring}\ and\ \citenamefont {Schuck}(1980)}]{RingSchuck}%
  \BibitemOpen
  \bibfield  {author} {\bibinfo {author} {\bibfnamefont {P.}~\bibnamefont {Ring}}\ and\ \bibinfo {author} {\bibfnamefont {P.}~\bibnamefont {Schuck}},\ }\href {\doibase 10.1063/1.2915762} {\emph {\bibinfo {title} {The Nuclear Many-Body Problems}}},\ Vol.\ \bibinfo {volume} {103}\ (\bibinfo {year} {1980})\BibitemShut {NoStop}%
\bibitem [{\citenamefont {Paar}\ \emph {et~al.}(2006)\citenamefont {Paar}, \citenamefont {Papakonstantinou}, \citenamefont {Hergert},\ and\ \citenamefont {Roth}}]{Paar2006}%
  \BibitemOpen
  \bibfield  {author} {\bibinfo {author} {\bibfnamefont {N.}~\bibnamefont {Paar}}, \bibinfo {author} {\bibfnamefont {P.}~\bibnamefont {Papakonstantinou}}, \bibinfo {author} {\bibfnamefont {H.}~\bibnamefont {Hergert}}, \ and\ \bibinfo {author} {\bibfnamefont {R.}~\bibnamefont {Roth}},\ }\href {\doibase 10.1103/PhysRevC.74.014318} {\bibfield  {journal} {\bibinfo  {journal} {Phys. Rev. C}\ }\textbf {\bibinfo {volume} {74}},\ \bibinfo {pages} {014318} (\bibinfo {year} {2006})}\BibitemShut {NoStop}%
\bibitem [{\citenamefont {Martini}\ \emph {et~al.}(2016)\citenamefont {Martini}, \citenamefont {P\'eru}, \citenamefont {Hilaire}, \citenamefont {Goriely},\ and\ \citenamefont {Lechaftois}}]{Martini2016}%
  \BibitemOpen
  \bibfield  {author} {\bibinfo {author} {\bibfnamefont {M.}~\bibnamefont {Martini}}, \bibinfo {author} {\bibfnamefont {S.}~\bibnamefont {P\'eru}}, \bibinfo {author} {\bibfnamefont {S.}~\bibnamefont {Hilaire}}, \bibinfo {author} {\bibfnamefont {S.}~\bibnamefont {Goriely}}, \ and\ \bibinfo {author} {\bibfnamefont {F.}~\bibnamefont {Lechaftois}},\ }\href {\doibase 10.1103/PhysRevC.94.014304} {\bibfield  {journal} {\bibinfo  {journal} {Phys. Rev. C}\ }\textbf {\bibinfo {volume} {94}},\ \bibinfo {pages} {014304} (\bibinfo {year} {2016})}\BibitemShut {NoStop}%
\bibitem [{\citenamefont {Goriely}\ \emph {et~al.}(2016)\citenamefont {Goriely}, \citenamefont {Hilaire}, \citenamefont {P\'eru}, \citenamefont {Martini}, \citenamefont {Deloncle},\ and\ \citenamefont {Lechaftois}}]{Goriely2016}%
  \BibitemOpen
  \bibfield  {author} {\bibinfo {author} {\bibfnamefont {S.}~\bibnamefont {Goriely}}, \bibinfo {author} {\bibfnamefont {S.}~\bibnamefont {Hilaire}}, \bibinfo {author} {\bibfnamefont {S.}~\bibnamefont {P\'eru}}, \bibinfo {author} {\bibfnamefont {M.}~\bibnamefont {Martini}}, \bibinfo {author} {\bibfnamefont {I.}~\bibnamefont {Deloncle}}, \ and\ \bibinfo {author} {\bibfnamefont {F.}~\bibnamefont {Lechaftois}},\ }\href {\doibase 10.1103/PhysRevC.94.044306} {\bibfield  {journal} {\bibinfo  {journal} {Phys. Rev. C}\ }\textbf {\bibinfo {volume} {94}},\ \bibinfo {pages} {044306} (\bibinfo {year} {2016})}\BibitemShut {NoStop}%
\bibitem [{\citenamefont {Goriely}\ \emph {et~al.}(2018)\citenamefont {Goriely}, \citenamefont {Hilaire}, \citenamefont {P\'eru},\ and\ \citenamefont {Sieja}}]{Goriely2018}%
  \BibitemOpen
  \bibfield  {author} {\bibinfo {author} {\bibfnamefont {S.}~\bibnamefont {Goriely}}, \bibinfo {author} {\bibfnamefont {S.}~\bibnamefont {Hilaire}}, \bibinfo {author} {\bibfnamefont {S.}~\bibnamefont {P\'eru}}, \ and\ \bibinfo {author} {\bibfnamefont {K.}~\bibnamefont {Sieja}},\ }\href {\doibase 10.1103/PhysRevC.98.014327} {\bibfield  {journal} {\bibinfo  {journal} {Phys. Rev. C}\ }\textbf {\bibinfo {volume} {98}},\ \bibinfo {pages} {014327} (\bibinfo {year} {2018})}\BibitemShut {NoStop}%
\bibitem [{\citenamefont {Porro}\ \emph {et~al.}(2024{\natexlab{a}})\citenamefont {Porro}, \citenamefont {Col{\`o}}, \citenamefont {Duguet}, \citenamefont {Gambacurta},\ and\ \citenamefont {Som{\`a}}}]{Porro:2023yto}%
  \BibitemOpen
  \bibfield  {author} {\bibinfo {author} {\bibfnamefont {A.}~\bibnamefont {Porro}}, \bibinfo {author} {\bibfnamefont {G.}~\bibnamefont {Col{\`o}}}, \bibinfo {author} {\bibfnamefont {T.}~\bibnamefont {Duguet}}, \bibinfo {author} {\bibfnamefont {D.}~\bibnamefont {Gambacurta}}, \ and\ \bibinfo {author} {\bibfnamefont {V.}~\bibnamefont {Som{\`a}}},\ }\href {\doibase 10.1103/PhysRevC.109.044315} {\bibfield  {journal} {\bibinfo  {journal} {Phys. Rev. C}\ }\textbf {\bibinfo {volume} {109}},\ \bibinfo {pages} {044315} (\bibinfo {year} {2024}{\natexlab{a}})},\ \Eprint {http://arxiv.org/abs/2312.10410} {arXiv:2312.10410 [nucl-th]} \BibitemShut {NoStop}%
\bibitem [{\citenamefont {Frosini}\ \emph {et~al.}(2024)\citenamefont {Frosini}, \citenamefont {Ryssens},\ and\ \citenamefont {Sieja}}]{Frosini2023}%
  \BibitemOpen
  \bibfield  {author} {\bibinfo {author} {\bibfnamefont {M.}~\bibnamefont {Frosini}}, \bibinfo {author} {\bibfnamefont {W.}~\bibnamefont {Ryssens}}, \ and\ \bibinfo {author} {\bibfnamefont {K.}~\bibnamefont {Sieja}},\ }\href {\doibase 10.1103/PhysRevC.110.014307} {\bibfield  {journal} {\bibinfo  {journal} {Phys. Rev. C}\ }\textbf {\bibinfo {volume} {110}},\ \bibinfo {pages} {014307} (\bibinfo {year} {2024})}\BibitemShut {NoStop}%
\bibitem [{\citenamefont {Bender}\ and\ \citenamefont {Heenen}(2008)}]{Bender08}%
  \BibitemOpen
  \bibfield  {author} {\bibinfo {author} {\bibfnamefont {M.}~\bibnamefont {Bender}}\ and\ \bibinfo {author} {\bibfnamefont {P.-H.}\ \bibnamefont {Heenen}},\ }\href {\doibase 10.1103/PhysRevC.78.024309} {\bibfield  {journal} {\bibinfo  {journal} {Phys. Rev. C}\ }\textbf {\bibinfo {volume} {78}},\ \bibinfo {pages} {024309} (\bibinfo {year} {2008})}\BibitemShut {NoStop}%
\bibitem [{\citenamefont {Niksic}\ \emph {et~al.}(2011)\citenamefont {Niksic}, \citenamefont {Vretenar},\ and\ \citenamefont {Ring}}]{Niksic11}%
  \BibitemOpen
  \bibfield  {author} {\bibinfo {author} {\bibfnamefont {T.}~\bibnamefont {Niksic}}, \bibinfo {author} {\bibfnamefont {D.}~\bibnamefont {Vretenar}}, \ and\ \bibinfo {author} {\bibfnamefont {P.}~\bibnamefont {Ring}},\ }\href {\doibase https://doi.org/10.1016/j.ppnp.2011.01.055} {\bibfield  {journal} {\bibinfo  {journal} {Progress in Particle and Nuclear Physics}\ }\textbf {\bibinfo {volume} {66}},\ \bibinfo {pages} {519} (\bibinfo {year} {2011})}\BibitemShut {NoStop}%
\bibitem [{\citenamefont {Robledo}\ \emph {et~al.}(2018)\citenamefont {Robledo}, \citenamefont {Rodríguez},\ and\ \citenamefont {Rodríguez-Guzmán}}]{Robledo19}%
  \BibitemOpen
  \bibfield  {author} {\bibinfo {author} {\bibfnamefont {L.~M.}\ \bibnamefont {Robledo}}, \bibinfo {author} {\bibfnamefont {T.~R.}\ \bibnamefont {Rodríguez}}, \ and\ \bibinfo {author} {\bibfnamefont {R.~R.}\ \bibnamefont {Rodríguez-Guzmán}},\ }\href {\doibase 10.1088/1361-6471/aadebd} {\bibfield  {journal} {\bibinfo  {journal} {Journal of Physics G: Nuclear and Particle Physics}\ }\textbf {\bibinfo {volume} {46}},\ \bibinfo {pages} {013001} (\bibinfo {year} {2018})}\BibitemShut {NoStop}%
\bibitem [{\citenamefont {Gao}\ \emph {et~al.}(2015)\citenamefont {Gao}, \citenamefont {Horoi},\ and\ \citenamefont {Chen}}]{Gao15}%
  \BibitemOpen
  \bibfield  {author} {\bibinfo {author} {\bibfnamefont {Z.-C.}\ \bibnamefont {Gao}}, \bibinfo {author} {\bibfnamefont {M.}~\bibnamefont {Horoi}}, \ and\ \bibinfo {author} {\bibfnamefont {Y.~S.}\ \bibnamefont {Chen}},\ }\href {\doibase 10.1103/PhysRevC.92.064310} {\bibfield  {journal} {\bibinfo  {journal} {Phys. Rev. C}\ }\textbf {\bibinfo {volume} {92}},\ \bibinfo {pages} {064310} (\bibinfo {year} {2015})}\BibitemShut {NoStop}%
\bibitem [{\citenamefont {Jiao}\ \emph {et~al.}(2017)\citenamefont {Jiao}, \citenamefont {Engel},\ and\ \citenamefont {Holt}}]{Jiao17}%
  \BibitemOpen
  \bibfield  {author} {\bibinfo {author} {\bibfnamefont {C.~F.}\ \bibnamefont {Jiao}}, \bibinfo {author} {\bibfnamefont {J.}~\bibnamefont {Engel}}, \ and\ \bibinfo {author} {\bibfnamefont {J.~D.}\ \bibnamefont {Holt}},\ }\href {\doibase 10.1103/PhysRevC.96.054310} {\bibfield  {journal} {\bibinfo  {journal} {Phys. Rev. C}\ }\textbf {\bibinfo {volume} {96}},\ \bibinfo {pages} {054310} (\bibinfo {year} {2017})}\BibitemShut {NoStop}%
\bibitem [{\citenamefont {Bally}\ \emph {et~al.}(2019)\citenamefont {Bally}, \citenamefont {S\'anchez-Fern\'andez},\ and\ \citenamefont {Rodr\'{\i}guez}}]{Bally19}%
  \BibitemOpen
  \bibfield  {author} {\bibinfo {author} {\bibfnamefont {B.}~\bibnamefont {Bally}}, \bibinfo {author} {\bibfnamefont {A.}~\bibnamefont {S\'anchez-Fern\'andez}}, \ and\ \bibinfo {author} {\bibfnamefont {T.~R.}\ \bibnamefont {Rodr\'{\i}guez}},\ }\href {\doibase 10.1103/PhysRevC.100.044308} {\bibfield  {journal} {\bibinfo  {journal} {Phys. Rev. C}\ }\textbf {\bibinfo {volume} {100}},\ \bibinfo {pages} {044308} (\bibinfo {year} {2019})}\BibitemShut {NoStop}%
\bibitem [{\citenamefont {S\'anchez-Fern\'andez}\ \emph {et~al.}(2021)\citenamefont {S\'anchez-Fern\'andez}, \citenamefont {Bally},\ and\ \citenamefont {Rodr\'{\i}guez}}]{Sanchez2021a}%
  \BibitemOpen
  \bibfield  {author} {\bibinfo {author} {\bibfnamefont {A.}~\bibnamefont {S\'anchez-Fern\'andez}}, \bibinfo {author} {\bibfnamefont {B.}~\bibnamefont {Bally}}, \ and\ \bibinfo {author} {\bibfnamefont {T.~R.}\ \bibnamefont {Rodr\'{\i}guez}},\ }\href {\doibase 10.1103/PhysRevC.104.054306} {\bibfield  {journal} {\bibinfo  {journal} {Phys. Rev. C}\ }\textbf {\bibinfo {volume} {104}},\ \bibinfo {pages} {054306} (\bibinfo {year} {2021})}\BibitemShut {NoStop}%
\bibitem [{\citenamefont {Dao}\ and\ \citenamefont {Nowacki}(2022)}]{Dao22}%
  \BibitemOpen
  \bibfield  {author} {\bibinfo {author} {\bibfnamefont {D.~D.}\ \bibnamefont {Dao}}\ and\ \bibinfo {author} {\bibfnamefont {F.}~\bibnamefont {Nowacki}},\ }\href {\doibase 10.1103/PhysRevC.105.054314} {\bibfield  {journal} {\bibinfo  {journal} {Phys. Rev. C}\ }\textbf {\bibinfo {volume} {105}},\ \bibinfo {pages} {054314} (\bibinfo {year} {2022})}\BibitemShut {NoStop}%
\bibitem [{\citenamefont {Kimura}(2017)}]{Kimura2017}%
  \BibitemOpen
  \bibfield  {author} {\bibinfo {author} {\bibfnamefont {M.}~\bibnamefont {Kimura}},\ }\href {https://link.aps.org/doi/10.1103/PhysRevC.95.034331} {\bibfield  {journal} {\bibinfo  {journal} {Phys. Rev. C}\ }\textbf {\bibinfo {volume} {95}},\ \bibinfo {pages} {034331} (\bibinfo {year} {2017})}\BibitemShut {NoStop}%
\bibitem [{\citenamefont {Frosini}\ \emph {et~al.}(2022{\natexlab{a}})\citenamefont {Frosini}, \citenamefont {Duguet}, \citenamefont {Ebran},\ and\ \citenamefont {Som{\`a}}}]{Frosini:2021fjf}%
  \BibitemOpen
  \bibfield  {author} {\bibinfo {author} {\bibfnamefont {M.}~\bibnamefont {Frosini}}, \bibinfo {author} {\bibfnamefont {T.}~\bibnamefont {Duguet}}, \bibinfo {author} {\bibfnamefont {J.-P.}\ \bibnamefont {Ebran}}, \ and\ \bibinfo {author} {\bibfnamefont {V.}~\bibnamefont {Som{\`a}}},\ }\href {\doibase 10.1140/epja/s10050-022-00692-z} {\bibfield  {journal} {\bibinfo  {journal} {Eur. Phys. J. A}\ }\textbf {\bibinfo {volume} {58}},\ \bibinfo {pages} {62} (\bibinfo {year} {2022}{\natexlab{a}})},\ \Eprint {http://arxiv.org/abs/2110.15737} {arXiv:2110.15737 [nucl-th]} \BibitemShut {NoStop}%
\bibitem [{\citenamefont {Frosini}\ \emph {et~al.}(2022{\natexlab{b}})\citenamefont {Frosini}, \citenamefont {Duguet}, \citenamefont {Ebran}, \citenamefont {Bally}, \citenamefont {Mongelli}, \citenamefont {Rodr{\'\i}guez}, \citenamefont {Roth},\ and\ \citenamefont {Som{\`a}}}]{Frosini:2021sxj}%
  \BibitemOpen
  \bibfield  {author} {\bibinfo {author} {\bibfnamefont {M.}~\bibnamefont {Frosini}}, \bibinfo {author} {\bibfnamefont {T.}~\bibnamefont {Duguet}}, \bibinfo {author} {\bibfnamefont {J.-P.}\ \bibnamefont {Ebran}}, \bibinfo {author} {\bibfnamefont {B.}~\bibnamefont {Bally}}, \bibinfo {author} {\bibfnamefont {T.}~\bibnamefont {Mongelli}}, \bibinfo {author} {\bibfnamefont {T.~R.}\ \bibnamefont {Rodr{\'\i}guez}}, \bibinfo {author} {\bibfnamefont {R.}~\bibnamefont {Roth}}, \ and\ \bibinfo {author} {\bibfnamefont {V.}~\bibnamefont {Som{\`a}}},\ }\href {\doibase 10.1140/epja/s10050-022-00693-y} {\bibfield  {journal} {\bibinfo  {journal} {Eur. Phys. J. A}\ }\textbf {\bibinfo {volume} {58}},\ \bibinfo {pages} {63} (\bibinfo {year} {2022}{\natexlab{b}})},\ \Eprint {http://arxiv.org/abs/2111.00797} {arXiv:2111.00797 [nucl-th]} \BibitemShut {NoStop}%
\bibitem [{\citenamefont {Frosini}\ \emph {et~al.}(2022{\natexlab{c}})\citenamefont {Frosini}, \citenamefont {Duguet}, \citenamefont {Ebran}, \citenamefont {Bally}, \citenamefont {Hergert}, \citenamefont {Rodr{\'\i}guez}, \citenamefont {Roth}, \citenamefont {Yao},\ and\ \citenamefont {Som{\`a}}}]{Frosini:2021ddm}%
  \BibitemOpen
  \bibfield  {author} {\bibinfo {author} {\bibfnamefont {M.}~\bibnamefont {Frosini}}, \bibinfo {author} {\bibfnamefont {T.}~\bibnamefont {Duguet}}, \bibinfo {author} {\bibfnamefont {J.-P.}\ \bibnamefont {Ebran}}, \bibinfo {author} {\bibfnamefont {B.}~\bibnamefont {Bally}}, \bibinfo {author} {\bibfnamefont {H.}~\bibnamefont {Hergert}}, \bibinfo {author} {\bibfnamefont {T.~R.}\ \bibnamefont {Rodr{\'\i}guez}}, \bibinfo {author} {\bibfnamefont {R.}~\bibnamefont {Roth}}, \bibinfo {author} {\bibfnamefont {J.}~\bibnamefont {Yao}}, \ and\ \bibinfo {author} {\bibfnamefont {V.}~\bibnamefont {Som{\`a}}},\ }\href {\doibase 10.1140/epja/s10050-022-00694-x} {\bibfield  {journal} {\bibinfo  {journal} {Eur. Phys. J. A}\ }\textbf {\bibinfo {volume} {58}},\ \bibinfo {pages} {64} (\bibinfo {year} {2022}{\natexlab{c}})},\ \Eprint {http://arxiv.org/abs/2111.01461} {arXiv:2111.01461 [nucl-th]} \BibitemShut {NoStop}%
\bibitem [{\citenamefont {Porro}\ \emph {et~al.}(2024{\natexlab{b}})\citenamefont {Porro}, \citenamefont {Duguet}, \citenamefont {Ebran}, \citenamefont {Frosini}, \citenamefont {Roth},\ and\ \citenamefont {Som{\'a}}}]{Porro:2024vlc}%
  \BibitemOpen
  \bibfield  {author} {\bibinfo {author} {\bibfnamefont {A.}~\bibnamefont {Porro}}, \bibinfo {author} {\bibfnamefont {T.}~\bibnamefont {Duguet}}, \bibinfo {author} {\bibfnamefont {J.-P.}\ \bibnamefont {Ebran}}, \bibinfo {author} {\bibfnamefont {M.}~\bibnamefont {Frosini}}, \bibinfo {author} {\bibfnamefont {R.}~\bibnamefont {Roth}}, \ and\ \bibinfo {author} {\bibfnamefont {V.}~\bibnamefont {Som{\'a}}},\ }\href {\doibase 10.1140/epja/s10050-024-01340-4} {\bibfield  {journal} {\bibinfo  {journal} {Eur. Phys. J. A}\ }\textbf {\bibinfo {volume} {60}},\ \bibinfo {pages} {133} (\bibinfo {year} {2024}{\natexlab{b}})},\ \Eprint {http://arxiv.org/abs/2402.02228} {arXiv:2402.02228 [nucl-th]} \BibitemShut {NoStop}%
\bibitem [{\citenamefont {Porro}\ \emph {et~al.}(2024{\natexlab{c}})\citenamefont {Porro}, \citenamefont {Duguet}, \citenamefont {Ebran}, \citenamefont {Frosini}, \citenamefont {Roth},\ and\ \citenamefont {Som{\`a}}}]{Porro:2024tzt}%
  \BibitemOpen
  \bibfield  {author} {\bibinfo {author} {\bibfnamefont {A.}~\bibnamefont {Porro}}, \bibinfo {author} {\bibfnamefont {T.}~\bibnamefont {Duguet}}, \bibinfo {author} {\bibfnamefont {J.-P.}\ \bibnamefont {Ebran}}, \bibinfo {author} {\bibfnamefont {M.}~\bibnamefont {Frosini}}, \bibinfo {author} {\bibfnamefont {R.}~\bibnamefont {Roth}}, \ and\ \bibinfo {author} {\bibfnamefont {V.}~\bibnamefont {Som{\`a}}},\ }\href {\doibase 10.1140/epja/s10050-024-01341-3} {\bibfield  {journal} {\bibinfo  {journal} {Eur. Phys. J. A}\ }\textbf {\bibinfo {volume} {60}},\ \bibinfo {pages} {134} (\bibinfo {year} {2024}{\natexlab{c}})},\ \Eprint {http://arxiv.org/abs/2402.15901} {arXiv:2402.15901 [nucl-th]} \BibitemShut {NoStop}%
\bibitem [{\citenamefont {Porro}\ \emph {et~al.}(2024{\natexlab{d}})\citenamefont {Porro}, \citenamefont {Duguet}, \citenamefont {Ebran}, \citenamefont {Frosini}, \citenamefont {Roth},\ and\ \citenamefont {Som{\`a}}}]{Porro:2024pdn}%
  \BibitemOpen
  \bibfield  {author} {\bibinfo {author} {\bibfnamefont {A.}~\bibnamefont {Porro}}, \bibinfo {author} {\bibfnamefont {T.}~\bibnamefont {Duguet}}, \bibinfo {author} {\bibfnamefont {J.-P.}\ \bibnamefont {Ebran}}, \bibinfo {author} {\bibfnamefont {M.}~\bibnamefont {Frosini}}, \bibinfo {author} {\bibfnamefont {R.}~\bibnamefont {Roth}}, \ and\ \bibinfo {author} {\bibfnamefont {V.}~\bibnamefont {Som{\`a}}},\ }\href {\doibase 10.1140/epja/s10050-024-01377-5} {\bibfield  {journal} {\bibinfo  {journal} {Eur. Phys. J. A}\ }\textbf {\bibinfo {volume} {60}},\ \bibinfo {pages} {155} (\bibinfo {year} {2024}{\natexlab{d}})},\ \Eprint {http://arxiv.org/abs/2404.14154} {arXiv:2404.14154 [nucl-th]} \BibitemShut {NoStop}%
\bibitem [{\citenamefont {Porro}\ \emph {et~al.}(2024{\natexlab{e}})\citenamefont {Porro}, \citenamefont {Duguet}, \citenamefont {Ebran}, \citenamefont {Frosini}, \citenamefont {Roth},\ and\ \citenamefont {Som{\`a}}}]{Porro:2024bid}%
  \BibitemOpen
  \bibfield  {author} {\bibinfo {author} {\bibfnamefont {A.}~\bibnamefont {Porro}}, \bibinfo {author} {\bibfnamefont {T.}~\bibnamefont {Duguet}}, \bibinfo {author} {\bibfnamefont {J.-P.}\ \bibnamefont {Ebran}}, \bibinfo {author} {\bibfnamefont {M.}~\bibnamefont {Frosini}}, \bibinfo {author} {\bibfnamefont {R.}~\bibnamefont {Roth}}, \ and\ \bibinfo {author} {\bibfnamefont {V.}~\bibnamefont {Som{\`a}}},\ }\href {\doibase 10.1140/epja/s10050-024-01448-7} {\bibfield  {journal} {\bibinfo  {journal} {Eur. Phys. J. A}\ }\textbf {\bibinfo {volume} {60}},\ \bibinfo {pages} {233} (\bibinfo {year} {2024}{\natexlab{e}})},\ \Eprint {http://arxiv.org/abs/2407.01325} {arXiv:2407.01325 [nucl-th]} \BibitemShut {NoStop}%
\bibitem [{\citenamefont {Brown}\ and\ \citenamefont {Richter}(2006)}]{USDB}%
  \BibitemOpen
  \bibfield  {author} {\bibinfo {author} {\bibfnamefont {B.~A.}\ \bibnamefont {Brown}}\ and\ \bibinfo {author} {\bibfnamefont {W.~A.}\ \bibnamefont {Richter}},\ }\href {\doibase 10.1103/PhysRevC.74.034315} {\bibfield  {journal} {\bibinfo  {journal} {Phys. Rev. C}\ }\textbf {\bibinfo {volume} {74}},\ \bibinfo {pages} {034315} (\bibinfo {year} {2006})}\BibitemShut {NoStop}%
\bibitem [{\citenamefont {Richter}\ \emph {et~al.}(2008)\citenamefont {Richter}, \citenamefont {Mkhize},\ and\ \citenamefont {Brown}}]{Richter2008}%
  \BibitemOpen
  \bibfield  {author} {\bibinfo {author} {\bibfnamefont {W.~A.}\ \bibnamefont {Richter}}, \bibinfo {author} {\bibfnamefont {S.}~\bibnamefont {Mkhize}}, \ and\ \bibinfo {author} {\bibfnamefont {B.~A.}\ \bibnamefont {Brown}},\ }\href {\doibase 10.1103/PhysRevC.78.064302} {\bibfield  {journal} {\bibinfo  {journal} {Phys. Rev. C}\ }\textbf {\bibinfo {volume} {78}},\ \bibinfo {pages} {064302} (\bibinfo {year} {2008})}\BibitemShut {NoStop}%
\bibitem [{\citenamefont {Richter}\ and\ \citenamefont {Brown}(2009)}]{Richter2009}%
  \BibitemOpen
  \bibfield  {author} {\bibinfo {author} {\bibfnamefont {W.~A.}\ \bibnamefont {Richter}}\ and\ \bibinfo {author} {\bibfnamefont {B.~A.}\ \bibnamefont {Brown}},\ }\href {\doibase 10.1103/PhysRevC.80.034301} {\bibfield  {journal} {\bibinfo  {journal} {Phys. Rev. C}\ }\textbf {\bibinfo {volume} {80}},\ \bibinfo {pages} {034301} (\bibinfo {year} {2009})}\BibitemShut {NoStop}%
\bibitem [{\citenamefont {Caurier}\ \emph {et~al.}(2005)\citenamefont {Caurier}, \citenamefont {Martinez-Pinedo}, \citenamefont {Nowacki}, \citenamefont {Poves},\ and\ \citenamefont {Zuker}}]{RMP}%
  \BibitemOpen
  \bibfield  {author} {\bibinfo {author} {\bibfnamefont {E.}~\bibnamefont {Caurier}}, \bibinfo {author} {\bibfnamefont {G.}~\bibnamefont {Martinez-Pinedo}}, \bibinfo {author} {\bibfnamefont {F.}~\bibnamefont {Nowacki}}, \bibinfo {author} {\bibfnamefont {A.}~\bibnamefont {Poves}}, \ and\ \bibinfo {author} {\bibfnamefont {A.~P.}\ \bibnamefont {Zuker}},\ }\href {\doibase 10.1103/RevModPhys.77.427} {\bibfield  {journal} {\bibinfo  {journal} {Rev. Mod. Phys.}\ }\textbf {\bibinfo {volume} {77}},\ \bibinfo {pages} {427} (\bibinfo {year} {2005})}\BibitemShut {NoStop}%
\bibitem [{\citenamefont {Schmid}(2004)}]{Schmid04}%
  \BibitemOpen
  \bibfield  {author} {\bibinfo {author} {\bibfnamefont {K.}~\bibnamefont {Schmid}},\ }\href {\doibase https://doi.org/10.1016/j.ppnp.2004.02.001} {\bibfield  {journal} {\bibinfo  {journal} {Prog. Part. Nucl. Phys.}\ }\textbf {\bibinfo {volume} {52}},\ \bibinfo {pages} {565} (\bibinfo {year} {2004})}\BibitemShut {NoStop}%
\bibitem [{\citenamefont {Shimizu}\ \emph {et~al.}(2021)\citenamefont {Shimizu}, \citenamefont {Tsunoda}, \citenamefont {Utsuno},\ and\ \citenamefont {T.}}]{Shimizu21}%
  \BibitemOpen
  \bibfield  {author} {\bibinfo {author} {\bibfnamefont {N.}~\bibnamefont {Shimizu}}, \bibinfo {author} {\bibfnamefont {Y.}~\bibnamefont {Tsunoda}}, \bibinfo {author} {\bibfnamefont {Y.}~\bibnamefont {Utsuno}}, \ and\ \bibinfo {author} {\bibfnamefont {O.}~\bibnamefont {T.}},\ }\href {\doibase 10.1103/PhysRevC.103.014312} {\bibfield  {journal} {\bibinfo  {journal} {Phys. Rev. C}\ }\textbf {\bibinfo {volume} {103}},\ \bibinfo {pages} {014312} (\bibinfo {year} {2021})}\BibitemShut {NoStop}%
\bibitem [{\citenamefont {Dao}\ and\ \citenamefont {Nowacki}(2025)}]{dao24}%
  \BibitemOpen
  \bibfield  {author} {\bibinfo {author} {\bibfnamefont {D.~D.}\ \bibnamefont {Dao}}\ and\ \bibinfo {author} {\bibfnamefont {F.}~\bibnamefont {Nowacki}},\ }\href {https://arxiv.org/abs/2409.08210} {\  (\bibinfo {year} {2025})},\ \Eprint {http://arxiv.org/abs/2409.08210} {arXiv:2409.08210 [nucl-th]} \BibitemShut {NoStop}%
\bibitem [{\citenamefont {Frosini}\ and\ \citenamefont {et~al.}()}]{panacea}%
  \BibitemOpen
  \bibfield  {author} {\bibinfo {author} {\bibfnamefont {M.}~\bibnamefont {Frosini}}\ and\ \bibinfo {author} {\bibnamefont {et~al.}},\ }\href@noop {} {\enquote {\bibinfo {title} {Pan@cea numerical suite},}\ }\BibitemShut {NoStop}%
\bibitem [{\citenamefont {Dufour}\ and\ \citenamefont {Zuker}(1996)}]{Dufour96}%
  \BibitemOpen
  \bibfield  {author} {\bibinfo {author} {\bibfnamefont {M.}~\bibnamefont {Dufour}}\ and\ \bibinfo {author} {\bibfnamefont {A.~P.}\ \bibnamefont {Zuker}},\ }\href {\doibase 10.1103/PhysRevC.54.1641} {\bibfield  {journal} {\bibinfo  {journal} {Phys. Rev. C}\ }\textbf {\bibinfo {volume} {54}},\ \bibinfo {pages} {1641} (\bibinfo {year} {1996})}\BibitemShut {NoStop}%
\bibitem [{\citenamefont {Bally}\ \emph {et~al.}(2021)\citenamefont {Bally}, \citenamefont {S\'anchez-Fernández},\ and\ \citenamefont {Rodr\'iguez}}]{Bally21}%
  \BibitemOpen
  \bibfield  {author} {\bibinfo {author} {\bibfnamefont {B.}~\bibnamefont {Bally}}, \bibinfo {author} {\bibfnamefont {A.}~\bibnamefont {S\'anchez-Fernández}}, \ and\ \bibinfo {author} {\bibfnamefont {T.~R.}\ \bibnamefont {Rodr\'iguez}},\ }\href {https://doi.org/10.1140/epja/s10050-021-00369-z} {\bibfield  {journal} {\bibinfo  {journal} {Eur. Phys. J. A}\ }\textbf {\bibinfo {volume} {57}},\ \bibinfo {pages} {69} (\bibinfo {year} {2021})}\BibitemShut {NoStop}%
\bibitem [{\citenamefont {Bally}\ and\ \citenamefont {Rodr\'iguez}(2024)}]{Bally24}%
  \BibitemOpen
  \bibfield  {author} {\bibinfo {author} {\bibfnamefont {B.}~\bibnamefont {Bally}}\ and\ \bibinfo {author} {\bibfnamefont {T.~R.}\ \bibnamefont {Rodr\'iguez}},\ }\href {https://doi.org/10.1140/epja/s10050-024-01271-0} {\bibfield  {journal} {\bibinfo  {journal} {Eur. Phys. J. A}\ }\textbf {\bibinfo {volume} {60}},\ \bibinfo {pages} {62} (\bibinfo {year} {2024})}\BibitemShut {NoStop}%
\bibitem [{\citenamefont {Mart\'inez-Larraz}\ and\ \citenamefont {Rodr\'iguez}(2022)}]{Martinez-Larraz22}%
  \BibitemOpen
  \bibfield  {author} {\bibinfo {author} {\bibfnamefont {J.}~\bibnamefont {Mart\'inez-Larraz}}\ and\ \bibinfo {author} {\bibfnamefont {T.~R.}\ \bibnamefont {Rodr\'iguez}},\ }\href {\doibase 10.1103/PhysRevC.106.054301} {\bibfield  {journal} {\bibinfo  {journal} {Phys. Rev. C}\ }\textbf {\bibinfo {volume} {106}},\ \bibinfo {pages} {054301} (\bibinfo {year} {2022})}\BibitemShut {NoStop}%
\bibitem [{NND()}]{NNDC}%
  \BibitemOpen
  \href@noop {} {}\bibinfo {howpublished} {http://www.nndc.bnl.gov/}\BibitemShut {NoStop}%
\bibitem [{\citenamefont {Schwengner}\ \emph {et~al.}(2013)\citenamefont {Schwengner}, \citenamefont {Frauendorf},\ and\ \citenamefont {Larsen}}]{Schwengner-Mo}%
  \BibitemOpen
  \bibfield  {author} {\bibinfo {author} {\bibfnamefont {R.}~\bibnamefont {Schwengner}}, \bibinfo {author} {\bibfnamefont {S.}~\bibnamefont {Frauendorf}}, \ and\ \bibinfo {author} {\bibfnamefont {A.~C.}\ \bibnamefont {Larsen}},\ }\href {https://link.aps.org/doi/10.1103/PhysRevLett.111.232504} {\bibfield  {journal} {\bibinfo  {journal} {Phys. Rev. Lett.}\ }\textbf {\bibinfo {volume} {111}},\ \bibinfo {pages} {232504} (\bibinfo {year} {2013})}\BibitemShut {NoStop}%
\bibitem [{\citenamefont {Brown}\ and\ \citenamefont {Larsen}(2014)}]{Brown-fe56}%
  \BibitemOpen
  \bibfield  {author} {\bibinfo {author} {\bibfnamefont {B.~A.}\ \bibnamefont {Brown}}\ and\ \bibinfo {author} {\bibfnamefont {A.~C.}\ \bibnamefont {Larsen}},\ }\href {\doibase 10.1103/PhysRevLett.113.252502} {\bibfield  {journal} {\bibinfo  {journal} {Phys. Rev. Lett.}\ }\textbf {\bibinfo {volume} {113}},\ \bibinfo {pages} {252502} (\bibinfo {year} {2014})}\BibitemShut {NoStop}%
\bibitem [{\citenamefont {Sieja}(2017)}]{Sieja-PRL}%
  \BibitemOpen
  \bibfield  {author} {\bibinfo {author} {\bibfnamefont {K.}~\bibnamefont {Sieja}},\ }\href {https://link.aps.org/doi/10.1103/PhysRevLett.119.052502} {\bibfield  {journal} {\bibinfo  {journal} {Phys. Rev. Lett.}\ }\textbf {\bibinfo {volume} {119}},\ \bibinfo {pages} {052502} (\bibinfo {year} {2017})}\BibitemShut {NoStop}%
\bibitem [{\citenamefont {Sieja}(2018)}]{Sieja2018}%
  \BibitemOpen
  \bibfield  {author} {\bibinfo {author} {\bibfnamefont {K.}~\bibnamefont {Sieja}},\ }\href {\doibase 10.1103/PhysRevC.98.064312} {\bibfield  {journal} {\bibinfo  {journal} {Phys. Rev. C}\ }\textbf {\bibinfo {volume} {98}},\ \bibinfo {pages} {064312} (\bibinfo {year} {2018})}\BibitemShut {NoStop}%
\end{thebibliography}%

\end{document}